\documentclass[final,3p,11pt]{elsarticle}

\usepackage{amsmath,amssymb,
latexsym,axodraw}
\usepackage{graphicx,xcolor}
\usepackage{times}
\usepackage{amsmath}
\usepackage{listings}
\usepackage{booktabs,tabularx}
\usepackage[bookmarks=true,linktocpage]{hyperref}

\hypersetup{pdftitle={Precise MSSM prediction for (g-2) of the muon},pdfauthor={P. Athron, M. Bach, H. G. Fargnoli, C. Gnendiger, R. Greifenhagen, J.-h.\ Park, S. Pa{\ss}ehr, D. St\"ockinger, H. St\"ockinger-Kim, A. Voigt},pdfkeywords={Anomalous,magnetic,moment,muon,supersymmetry,MSSM}}
\usepackage[absolute]{textpos}

\definecolor{light}{RGB}{245,245,245}

\lstdefinestyle{Mybash}{
  language=bash,
  backgroundcolor=\color{light}
}

\bibstyle{elsarticle-num}
\biboptions{numbers,sort&compress}

\def\amu{a_\mu}

\newcommand{\amuSUOL}{a_\mu^{\rm 1L}}
\newcommand{\amuFSf}{a_\mu^{{\rm 2L,} f\tilde{f}}}
\newcommand{\amuTLa}{a_\mu^{\rm 2L(a)}}
\newcommand{\amuphotonic}{a_\mu^{\rm 2L,\ photonic}}
\newcommand{\MSbar}{\ensuremath{\overline{\text{MS}}}}
\newcommand{\DRbar}{$\overline{\text{DR}}$}
\newcommand{\OURPROGRAM}{GM2Calc}
\newcommand{\SOFTSUSY}{SOFTSUSY}
\newcommand{\SPHENO}{SPheno}
\newcommand{\code}[1]{\lstinline|#1|}  
\newcommand{\unit}[1]{\,\text{#1}}

\begin{document}
\begin{flushright}
\end{flushright}
\title{\Large\bf \OURPROGRAM: Precise MSSM prediction for \boldmath$(g-2)$ of
  the muon}
\author{Peter Athron$^{a}$, Markus Bach$^b$, Helvecio G.\ Fargnoli$^{c}$, Christoph Gnendiger$^b$,\\
  Robert Greifenhagen$^b$,
  Jae-hyeon Park$^d$, Sebastian Pa{\ss}ehr$^e$, Dominik St\"ockinger$^b$,\\
Hyejung St\"ockinger-Kim$^b$, Alexander Voigt$^e$ \vspace{1em}
}

\address{\it $^a$ARC Centre of Excellence for Particle Physics at 
the Terascale, School of Physics, Monash University, Victoria 3800}
\address{\it ${}^b$Institut f\"ur Kern- und Teilchenphysik,
TU Dresden, Zellescher Weg 19, 01069 Dresden, Germany}
\address{\it ${}^c$Departamento de Ci\^encias Exatas, Universidade Federal de Lavras, 37200-000, Lavras, Brazil}
\address{\it $^d$Departament de F\'{i}sica Te\`{o}rica and IFIC,
Universitat de Val\`{e}ncia-CSIC,
46100, Burjassot, Spain}
\address{\it ${}^e$Deutsches Elektronen-Synchrotron DESY,
  Notkestra\ss e 85, 22607 Hamburg, Germany}
\setcounter{footnote}{0}
\begin{abstract}
  We present \OURPROGRAM, a public C++ program for the calculation of
  MSSM contributions to the anomalous magnetic moment of the muon,
  $(g-2)_\mu$. The code computes $(g-2)_\mu$ precisely, by taking into
  account the latest two-loop corrections and by performing the calculation
  in a physical on-shell renormalization scheme.  In particular the program
  includes a $\tan\beta$ resummation so that it is valid for arbitrarily high
  values of $\tan\beta$, as well as fermion/sfermion-loop corrections which
  lead to non-decoupling effects from heavy squarks. \OURPROGRAM\ can be
  run with a standard SLHA input file, internally converting the input into on-shell
  parameters. Alternatively, input parameters may be specified directly in
  this on-shell scheme. In both cases the input file allows one to switch
  on/off individual contributions to study their relative impact. This paper
  also provides typical usage examples not only in conjunction with
  spectrum generators and plotting programs but also as C++ subroutines
  linked to other programs.
\end{abstract}

\maketitle 

\begin{textblock*}{7em}(\textwidth,1cm)
\noindent\footnotesize
CoEPP--MN--15--10 \\
DESY 15-193 \\
FTUV--15--6502 \\
IFIC--15--76
\end{textblock*}

\section{Introduction}

Important constraints on physics beyond the Standard Model (SM) are
provided by the anomalous magnetic moment of the muon
$a_\mu=(g-2)_\mu/2$. The SM prediction and the experimental
determination~\cite{Bennett:2006fi} have both reached sub-ppm
precision, and there is a tantalizing deviation of more than 3 standard
deviations,
\begin{align}
a_\mu^{\text{Exp$-$SM}}&= 
\begin{cases}
(28.7 \pm 8.0 ) \times 10^{-10} \mbox{\cite{Davier}}, \\
(26.1 \pm 8.0 ) \times 10^{-10} \mbox{\cite{HMNT}},
\end{cases}
\label{deviation}
\end{align}
using the indicated references for the hadronic vacuum polarization
contributions.\footnote{The numbers take into account the most recent
  refinements on the QED~\cite{Kinoshita2012} and
  electroweak~\cite{Gnendiger:2013pva} contributions. For further
  recent theoretical progress on QED and hadronic contributions and
  reviews, see Refs.~\cite{Kataev:2012kn,SteinhauserQED}, 
  \cite{JegerlehnerSzafron,Benayoun:2012wc,Kurz:2014wya,Colangelo:2014qya,Colangelo:2014,Pauk:2014rfa,Blum,Ablikim:2015orh}, 
  and~\cite{JegerlehnerNyffeler,Miller:2012opa,Blum:2013xva,Benayoun:2014tra}, 
  respectively.}  Importantly, a fourfold improvement in precision is
expected from the new experiments at Fermilab and
J-PARC~\cite{Carey:2009zzb,Iinuma:2011zz}, which promises to further
strengthen the power of $\amu$ to constrain and identify new
physics~\cite{WhitePaper,Miller:2012opa}.

The minimal supersymmetric standard model (MSSM) is one of the best
motivated and most studied extensions of the SM\@. It could also provide
a promising explanation of the deviation \eqref{deviation}, as
reviewed in
Refs.~\cite{CzM,MartinWells,DSreview,Cho:2011rk}. Therefore, $\amu$
has been employed extensively as a constraint on the MSSM parameter
space. Recent studies, which also focus on the complementarity to and
correlations with other observables, are discussed in
Refs.~\cite{Endo,Evans:2012hg,Ibe:2013oha,Akula:2013ioa,Bhattacharyya:2013xma,Mohanty:2013soa,Gogoladze:2014cha,Kersten:2014xaa,Chiu:2014oma,Badziak:2014kea,Kowalska:2015zja,Wang:2015rli,Calibbi:2015kja,Okada:2013ija,Li:2014dna}.
The importance and the predictivity of the MSSM have led to the
development of many advanced computer codes. The set of available
programs ranges from spectrum generators
(\SOFTSUSY~\cite{Allanach:2001kg}, \SPHENO~\cite{spheno},
Suspect~\cite{Djouadi:2002ze}, IsaSusy~\cite{Baer:1993ae},
SUSEFLAV~\cite{Chowdhury:2011zr}, SARAH/\SPHENO~\cite{Sarah},
FlexibleSUSY~\cite{Athron:2014yba})
  to calculators for specific observables. Codes which
include the computation of $a_\mu$ in the MSSM are
SuperISO~\cite{Mahmoudi:2008tp}, FeynHiggs~\cite{Hahn:2009zz},
SusyFlavor~\cite{Rosiek:2010ug}, and CPSuperH~\cite{Lee:2012wa}. The
SUSY Les Houches accord version~1\ \cite{Skands:2003cj} (SLHA) has been
established as an efficient standard for passing information between
programs.

Here we present \OURPROGRAM, a C++ program that calculates $\amu$ in
the MSSM.\footnote{At the moment, only the lepton-flavour conserving
  case is implemented. It is planned to extend the program to the
  lepton-flavour violating case in the future.}  It can be run with
SLHA input or with a \OURPROGRAM-specific input file, and it computes
$\amu$ in the MSSM fast and precisely, taking into account the
recently computed 
two-loop contributions. In particular, in contrast to the existing
public codes, it contains the fermion/sfermion-two-loop
corrections~\cite{Fargnoli:2013zda,Fargnoli:2013zia}, which include
the universal correction $\Delta\rho$ and potentially large
non-decoupling logarithms of heavy squark or slepton masses. \OURPROGRAM\ 
further provides resummation of $n$-loop $(\tan\beta)^n$
contributions~\cite{Marchetti:2008hw,Bach:2015doa}, which allows for
arbitrarily high $\tan\beta$ ($\tan\beta$ being the ratio of the
vacuum expectation values of the two Higgs doublets). It goes beyond
Refs.~\cite{Fargnoli:2013zda,Fargnoli:2013zia,Marchetti:2008hw,Bach:2015doa}
in that it implements the $\tan\beta$-resummation also for the
two-loop contributions.

The code uses routines from FlexibleSUSY~\cite{Athron:2014yba}, but it
is a standalone code which does not require an installation of
FlexibleSUSY or any of its prerequisites such as SARAH~\cite{Sarah}.

In Section~\ref{sec:contributionsandinput} we describe the different
contributions to $a_\mu$, provide an estimate for the theory uncertainty,
and define the input parameters which are relevant for the computation.
Afterwards, an explanation of how to use and customize \OURPROGRAM\
is given in Section~\ref{sec:usage}, together with a description of the
different input and output formats.
Section~\ref{sec:examples} includes several examples for practical
applications of \OURPROGRAM, which can easily be extended and adapted
to perform sophisticated studies.
We summarize our explanations and provide some final comments in
Section~\ref{sec:summary}. The Appendix describes all implemented
formulas and lists sample input files.

\section{Implemented contributions and definition of input}
\label{sec:contributionsandinput}

The implementation of the SUSY contributions to $\amu$, i.\,e.\ the
difference between the full MSSM and the full SM prediction for $\amu$,
follows the decomposition introduced in Ref.~\cite{DSreview}. There,
the SUSY two-loop corrections are split into class 2L(a), which
corresponds to corrections to SM-like one-loop diagrams, and class
2L(b), which corresponds to corrections to SUSY one-loop diagrams. The
contributions of class 2L(b) are further subdivided. The
implementation of \OURPROGRAM\ can 
be written as
\begin{align}
\amu^{\text{SUSY}} &= \left[\amuSUOL  +\amuTLa
+
\amuphotonic 
 +\amuFSf\right]_{t_\beta\text{-resummed}}
+\cdots\ .
\label{amuSUSYdecomposition}
\end{align}
The dots represent further two-loop contributions of class 2L(b) and
higher-order contributions,  which are not known and which have
not been implemented into \OURPROGRAM.
In the following we briefly describe the individual contributions and
their phenomenological impact, and we provide an estimate of the
theory uncertainty. The implemented formulas can be
found in the appendix. Afterwards we describe the definitions of the input parameters and their renormalization scheme.

\subsection{Individual contributions to $\amu$}
\label{sec:contributions}

\begin{figure}
\null\hfill
\scalebox{.65}{\setlength{\unitlength}{1pt}
\begin{picture}(240,100)(-120,0)
\CArc(0,0)(60,0,180)
\ArrowLine(-120,0)(-60,0)
\ArrowLine(60,0)(120,0)
\DashArrowLine(-60,0)(60,0){4}
\Photon(-100,100)(-60,60){4}{5}
\Vertex(60,0){2}
\Vertex(-60,0){2}
\Text(-80,95)[]{\scalebox{1.35}{$\gamma$}}
\Text(-100,-10)[]{\scalebox{1.35}{$\mu$}}
\Text(100,-10)[]{\scalebox{1.35}{${\mu}$}}
\Text(0,-10)[c]{\scalebox{1.35}{$\tilde{\mu}, \tilde{\nu}_{\mu}$}}
\Text(0,50)[c]{\scalebox{1.35}{$\chi_{i} ^{0,-}$}}
\end{picture}}
\hfill\hfill
\scalebox{.65}{\setlength{\unitlength}{1pt}
\begin{picture}(240,100)(-120,0)
\PhotonArc(0,0)(60,0,60){4}{6}
\DashCArc(0,0)(60,120,180){4}
\DashArrowArcn(0,51)(30,180,0){4}
\DashArrowArcn(0,51)(30,0,180){4}
\ArrowLine(-120,0)(-60,0)
\ArrowLine(60,0)(120,0)
\ArrowLine(-60,0)(60,0)
\Photon(-100,100)(-60,60){4}{5}
\Vertex(30,51){2}
\Vertex(-30,51){2}
\Vertex(60,0){2}
\Vertex(-60,0){2}
\Text(-80,95)[]{\scalebox{1.35}{$\gamma$}}
\Text(-100,-8)[]{\scalebox{1.35}{$\mu$}}
\Text(100,-8)[]{\scalebox{1.35}{${\mu}$}}
\Text(0,-8)[c]{\scalebox{1.35}{$\mu$}}
\Text(-40,20)[]{\scalebox{1.35}{$H$}}
\Text(40,20)[]{\scalebox{1.35}{$\gamma$}}
\Text(0,70)[]{\scalebox{1.35}{$\tilde t$}}
\Text(0,32)[]{\scalebox{1.35}{$\tilde t$}}
\end{picture}}
\null\hfill
\null 
\caption{SUSY one-loop diagrams (left) and sample diagram of class 2L(a) with
  closed stop loop inserted into 
  an SM-like one-loop diagram with Higgs- and photon-exchange (right). The external
  photon can couple to each charged particle.}
\label{fig:diagramsA}
\end{figure}
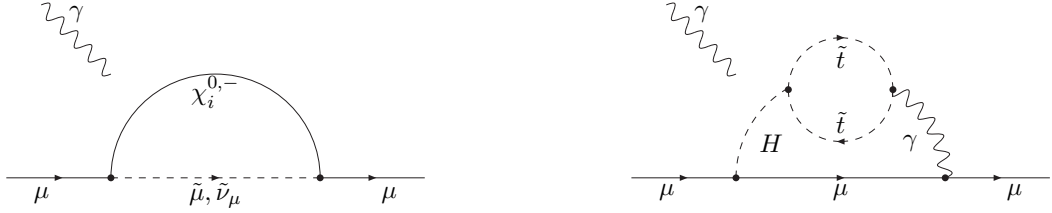

\begin{figure}
\begin{center}
\null\hfill
\scalebox{.65}{\setlength{\unitlength}{1pt}
\begin{picture}(240,100)(-120,0)
\CArc(0,0)(60,0,180)
\PhotonArc(0,0)(75,0,180){4}{16}
\ArrowLine(-120,0)(-60,0)
\ArrowLine(60,0)(120,0)
\DashArrowLine(-60,0)(60,0){4}
\Photon(-100,100)(-60,60){4}{5}
\Vertex(60,0){2}
\Vertex(-60,0){2}
\Text(-80,95)[]{\scalebox{1.35}{$\gamma$}}
\Text(62,62)[c]{\scalebox{1.35}{$\gamma$}}
\Text(-100,-10)[]{\scalebox{1.35}{$\mu$}}
\Text(100,-10)[]{\scalebox{1.35}{${\mu}$}}
\Text(0,-10)[c]{\scalebox{1.35}{$\tilde{\mu}, \tilde{\nu}_{\mu}$}}
\Text(0,50)[c]{\scalebox{1.35}{$\chi_{i} ^{0,-}$}}
\end{picture}}
\hfill\hfill
\scalebox{.65}{\setlength{\unitlength}{1pt}
\begin{picture}(240,100)(-120,0)
\CArc(0,0)(60,0,60)
\CArc(0,0)(60,120,180)
\ArrowArcn(0,51)(30,180,0)
\DashArrowArcn(0,51)(30,0,180){4}
\ArrowLine(-120,0)(-60,0)
\ArrowLine(60,0)(120,0)
\DashArrowLine(-60,0)(60,0){4}
\Photon(-100,100)(-60,60){4}{5}
\Vertex(30,51){2}
\Vertex(-30,51){2}
\Vertex(60,0){2}
\Vertex(-60,0){2}
\Text(-80,95)[]{\scalebox{1.35}{$\gamma$}}
\Text(-100,-10)[]{\scalebox{1.35}{$\mu$}}
\Text(100,-10)[]{\scalebox{1.35}{${\mu}$}}
\Text(0,-10)[c]{\scalebox{1.35}{$\tilde{\mu}, \tilde{\nu}_{\mu}$}}
\Text(-40,20)[]{\scalebox{1.35}{$\chi_{j} ^{0,-}$}}
\Text(40,20)[]{\scalebox{1.35}{$\chi_{i} ^{0,-}$}}
\Text(0,70)[]{\scalebox{1.35}{$f,f'$}}
\Text(0,32)[]{\scalebox{1.35}{$\tilde f$}}
\end{picture}}
\hfill\null
\end{center}
\caption{\label{fig:diagramsB}
Sample two-loop diagrams corresponding to SUSY one-loop diagrams with
additional photon loop (left), or with
  fermion/sfermion-loop insertion (right). The
  external photon can couple to each charged particle.}
\end{figure}
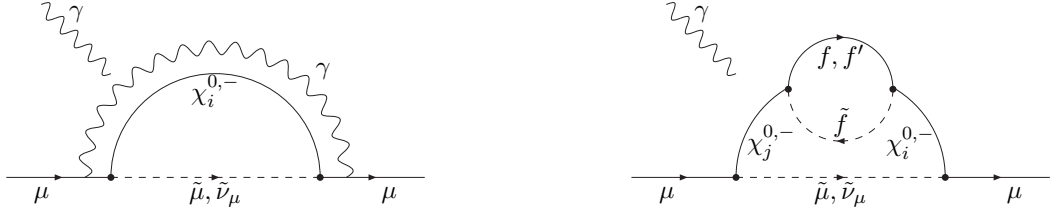
\begin{itemize}
\item One-loop corrections, $\amuSUOL$ (see
  Fig.~\ref{fig:diagramsA}(left)):\\
  The one-loop contributions arise from Feynman diagrams with the
  exchange of the SUSY partners of the muon or neutrino, smuon
  $\tilde{\mu}$ or sneutrino $\tilde{\nu}_\mu$, and the SUSY partners
  of the Higgs and gauge Bosons, the neutralinos and charginos
  $\chi^{0,\pm}$. They have
  been computed in full generality in Ref.~\cite{moroi}, see
  also Refs.~\cite{MartinWells,DSreview,Cho:2011rk} for reviews and
  discussions. 
These one-loop corrections depend essentially on the bino/wino masses $M_{1/2}$, the Higgsino mass $\mu$, the left- and right-smuon mass parameters, $M_{L2}, M_{E2}$, and the ratio of the two Higgs vacuum expectation values, $\tan\beta$. They have a weak dependence on the second generation $A$-parameter $A^e_{2,2}$,
and show a simple scaling behaviour
  $\propto\tan\beta/M_{\text{SUSY}}^2$, 
where $M_{\text{SUSY}}$ is a generic SUSY mass scale. However, the
  detailed dependence on the five relevant mass parameters is
  intricate. It can be understood particularly well with the help of
  mass-insertion diagrams, and it is possible to obtain large
  contributions even in presence of very heavy SUSY
  particles~\cite{moroi,Cho:2011rk,Fargnoli:2013zda}.
\item Two-loop corrections to SM-like one-loop diagrams, $\amuTLa$
  (see Fig.~\ref{fig:diagramsA}(right)):\\
  These class 2L(a) contributions are interesting 
since they do not depend on smuon masses but instead on Higgs Boson masses and on the squark and slepton masses of all generations.
They can be large in certain regions of parameter space, but they show decoupling behaviour and become small
  as the masses of SUSY particles or heavy Higgs Bosons become
  large. The exact results are reported in Refs.~\cite{HSW03,HSW04}; in \OURPROGRAM\
  we have implemented the good approximation in terms of photonic
  Barr-Zee diagrams~\cite{ArhribBaek,ChenGeng}, where a pure SUSY loop
  (of either charginos, neutralinos, or sfermions) is inserted into an
  effective Higgs--$\gamma$--$\gamma$ interaction in an SM-like
  diagram.  
\item MSSM photonic two-loop corrections, $\amuphotonic$  (see
  Fig.~\ref{fig:diagramsB}(left)):\\
  These corrections correspond to SUSY one-loop diagrams with an additional photon exchange; they have been
  evaluated in Ref.~\cite{vonWeitershausen:2010zr}. They include the
  large QED logarithm $\log(M_{\text{SUSY}}/m_\mu)$~\cite{DG98}, which has a negative prefactor and
  typically leads to a $(-7\ldots{-9})\%$ correction, and
  further terms which depend on the individual SUSY masses and which
  can be positive or negative.
\item MSSM fermion/sfermion-loop corrections, $\amuFSf$ (see
  Fig.~\ref{fig:diagramsB}(right)):\\
  The two-loop fermion/sfermion-loop contributions  presented in
  Refs.~\cite{Fargnoli:2013zda,Fargnoli:2013zia} introduce a
  dependence of~$a_\mu$ on squarks and sleptons of all generations,
  which is phenomenologically interesting. Most notably, if the squark
  masses (or slepton masses of the first or third generation) become
  large, the contributions to~$a_{\mu}$ do not decouple but are
  logarithmically
  enhanced. Depending on the mass pattern,
  Refs.~\cite{Fargnoli:2013zda,Fargnoli:2013zia} have found 
  positive or negative corrections of ${\cal O}(10\%)$ for squark masses in the
  few-TeV region. The fermion/sfermion-loop contributions further contain the universal
  quantity $\Delta\rho$ and significantly reduce the theory
  uncertainty arising from the possibility of different
  parametrizations of the fine-structure constant $\alpha$ in the
  one-loop contributions.
\item $\tan\beta$ resummation:\\
  The subscript $_{t_\beta\text{-resummed}}$ indicates that $n$-loop
  terms  $\propto\left(\tan\beta\right)^n$
  have been resummed to all
  orders. According to Refs.~\cite{Marchetti:2008hw,Bach:2015doa}, the
  resummation is carried out by evaluating the muon Yukawa coupling
  not at tree level, but in the form
\begin{align}
y_\mu &= \frac{m_{\mu}\,e}{\sqrt{2}\,s_W M_{W} \cos{\beta}\; (1 + \Delta_\mu)},
\label{ymuresummation}
\end{align}
where $e$ is the positron charge and $s_W=\sqrt{1-M_W^2/M_Z^2}$, and
where $\Delta_\mu$ contains $\tan\beta$-enhanced loop contributions
to the muon self energy. Analogous replacements are carried out
also for the third-generation down-type Yukawa couplings, $y_\tau$ and
$y_b$, which appear in the two-loop contributions.
For values of
$\tan\beta$ up to 50, these higher-order effects can amount to
corrections of up to $10\%$~\cite{Marchetti:2008hw}. Including these
resummations also allows setting $\tan\beta$ to an arbitrarily high value
which can be used to approximate the limit $\tan\beta\to\infty$. As
studied in Ref.~\cite{Bach:2015doa},\footnote{The viability of this
  limit from the point of view of $B$- and Higgs-physics has also been
  studied in Refs.~\cite{Dobrescu:2010mk,Altmannshofer:2010zt}.} this
limit has a distinctive phenomenology and allows for large SUSY
contributions to $\amu$ even if all SUSY masses are at or above the
TeV scale.
\end{itemize}
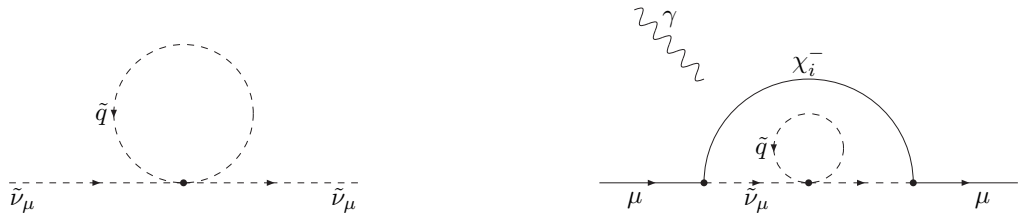
\begin{figure}
\null\hfill
\scalebox{.65}{\setlength{\unitlength}{1pt}
\begin{picture}(240,100)(-120,0)
\DashArrowLine(-100,0)(0,0){4}
\DashArrowLine(0,0)(100,0){4}
\DashArrowArc(0,40)(40,0,360){4}
\Text(-100,-10)[l]{\scalebox{1.35}{$\tilde{\nu}_\mu$}}
\Text(100,-10)[r]{\scalebox{1.35}{$\tilde{\nu}_\mu$}}
\Text(-40,40)[r]{\scalebox{1.35}{$\tilde{q}\ $}}
\Vertex(0,0){2}
\end{picture}}
\hfill\hfill
\scalebox{.65}{\setlength{\unitlength}{1pt}
\begin{picture}(240,100)(-120,0)
\CArc(0,0)(60,0,180)
\ArrowLine(-120,0)(-60,0)
\ArrowLine(60,0)(120,0)
\DashArrowLine(-60,0)(0,0){4}
\DashArrowLine(0,0)(60,0){4}
\DashArrowArc(0,20)(20,0,360){4}
\Photon(-100,100)(-60,60){4}{5}
\Vertex(60,0){2}
\Vertex(0,0){2}
\Vertex(-60,0){2}
\Text(-20,20)[r]{\scalebox{1.35}{$\tilde{q}\ $}}
\Text(-80,95)[]{\scalebox{1.35}{$\gamma$}}
\Text(-100,-10)[]{\scalebox{1.35}{$\mu$}}
\Text(100,-10)[]{\scalebox{1.35}{${\mu}$}}
\Text(-30,-10)[c]{\scalebox{1.35}{$\tilde{\nu}_{\mu}$}}
\Text(0,62)[b]{\scalebox{1.35}{$\chi_{i} ^{-}$}}
\end{picture}}
\hfill
\null
\caption{\label{fig:smuonSE}
Left: Sample muon sneutrino self energy diagram with squark loop, which gives rise to
corrections of order $m_{\tilde{q}}^2$ between the sneutrino pole and
\DRbar\ masses. Right: Two-loop diagram for
$\amu^{\text{SUSY}}$ with an insertion of the left diagram. }
\end{figure}

\subsection{Estimate of theory uncertainty}
\label{sec:error}

It is important to estimate the theory uncertainty due to missing
contributions. 
Ref.~\cite{DSreview} has given estimates for the still unknown two-loop contributions at the time, as well as for
the employed approximation for $\amuTLa$. Updating this estimate with the now known fermion/sfermion-loop and photonic two-loop corrections and the $\tan\beta$-resummation, we obtain
\begin{align}
\delta\amu^{\text{SUSY}} &=
2.3\times10^{-10}
+ 0.3\,
\left(|\amu^{(\chi\gamma H)}|+|\amu^{(\tilde{f}\gamma H)}|\right)
\,
.
\label{error}
\end{align}
We give the following comments on this error estimate:
\begin{itemize}
\item  The error is significantly smaller than the uncertainty of
  Eq.~\eqref{deviation}, but it will become critical once data from the
  improved Fermilab or J-PARC $(g-2)$ measurements is available.
\item The error estimate is deliberately conservative, see
  Ref.~\cite{DSreview}. To improve the precision reliably, however,
  the full two-loop computation of $\amu^\text{SUSY}$ will be
  necessary. 
\item The error estimate has been derived for our case of the on-shell
  renormalization scheme (see below). The difference between the
  result of \OURPROGRAM\ and evaluations/codes using e.\,g.\ the
  \DRbar\ scheme can be much larger than
  Eq.~\eqref{error}. Fig.~\ref{fig:smuonSE}(left) shows an example Feynman
  diagram which leads to differences of order
  $m_{\tilde{q}}^2$ between the muon sneutrino pole and
  \DRbar\ masses. The differences can be arbitrarily large for large
  squark masses $m_{\tilde{q}}$. Fig.~\ref{fig:smuonSE}(right) shows a corresponding two-loop
  contribution to $\amu^{\text{SUSY}}$ which is quadratically
  sensitive to $m_{\tilde{q}}$ in the
  \DRbar\ scheme. In the on-shell scheme used here
  these diagrams are cancelled by counterterms and thus the large
  contributions are avoided.
\end{itemize}

\subsection{Input parameters and renormalization scheme}
\label{sec:input_parameters}

The program can be run with two different input formats:
\begin{enumerate}
\item SLHA input format: contains the pole masses of the SM and SUSY
  particles as well as running \DRbar\ parameters. For this choice a
  detailed knowledge of the renormalization scheme is not required;
  the program does all renormalization scheme conversions internally
  and automatically. It should only be noted that the values of the
  gauge couplings are not taken from the SLHA file, but from a
  hardcoded value of the fine-structure constant $\alpha$ (this
  value can be overridden as described in
  Section~\ref{sec:usage}). Readers who are only interested in using
  SLHA input can skip the remaining section and continue reading in
  Section~\ref{sec:usage}, where the usage is explained in detail.
\item \OURPROGRAM-specific input format: specifies SM
  and MSSM parameters as defined in the mostly-on-shell
  renormalization scheme of
  Refs.~\cite{vonWeitershausen:2010zr,Fargnoli:2013zda,Fargnoli:2013zia}.
\end{enumerate}
In the following we describe in more detail the internally-used input
parameters and how they are obtained from the input files.  The
internal implementation uses the renormalization scheme of
Refs.~\cite{vonWeitershausen:2010zr,Fargnoli:2013zda,Fargnoli:2013zia},
which corresponds to on-shell renormalization of the MSSM as far as
possible, similarly to the schemes of
Refs.~\cite{HKRRSS,tf,Heidi,Heinemeyer:2010mm,Fritzsche:2011nr,HeinemeyerNew}.

The following parameters appear in the implementation of the one-loop
contributions $\amuSUOL$:
\begin{align}
\alpha(M_Z), M_{W,Z}, m_\mu; \quad
\tan\beta(Q);
\quad
 M_1, M_2, \mu,
M_{L2}, M_{E2};
\quad A^e_{2,2}(Q).
\end{align}
Here $M_{W,Z}$ and $m_\mu$ denote the SM masses of $W$, $Z$, and muon,
defined as pole masses in the on-shell scheme.  The fine-structure
constant is defined as $\alpha(M_{Z}) = {\alpha(0)}/{(1 - \Delta
  \alpha(M_{Z}))}$ where $\alpha(0)$ is the value in the Thomson limit
and $\Delta\alpha(M_Z)$ arises from quark and lepton contributions to
the on-shell renormalized photon vacuum polarization. Note that this
definition is different from the \MSbar\ or
\DRbar\ definitions, which would be provided by the
SLHA standard.

All these SM input parameters can be given explicitly, or they can be
omitted from the input files. In the latter case, hardcoded values are
used. For the pole masses, the hardcoded values are the current PDG
values~\cite{PDG}; in case of the fine-structure constant the
hardcoded value $\alpha(M_Z)=1/128.944$ based on Ref.~\cite{HMNT} is
used.

The ratio of the Higgs vacuum expectation values $\tan\beta$ is defined in the
\DRbar\ scheme~\cite{Freitas:2002um} at the scale~$Q$. The
\DRbar\ scheme at scale $Q$ is also chosen for the
trilinear soft breaking parameter $A^e_{2,2}$ entering in the smuon
mixing matrix.

The remaining five one-loop parameters are SUSY mass parameters,
defined in the on-shell scheme according to the following conditions:
in the chargino sector the wino and Higgsino masses $M_2$ and $\mu$
are chosen such that the two tree-level chargino masses coincide with
the corresponding pole masses. In the neutralino sector the bino mass
$M_1$ is defined by the requirement that the tree-level mass and the
pole mass of the bino-like neutralino coincide.  Similarly, the two
smuon mass parameters $M_{L2}$, $M_{E2}$ are chosen such that the
tree-level and pole masses of the muon sneutrino and the mostly
right-handed smuon coincide.

In case of the \OURPROGRAM\ input format these five SUSY mass
parameters are provided directly in the on-shell renormalization
scheme, and no internal conversion is carried out. In case of the SLHA
input format, the relevant information is provided by the pole masses
of the charginos, the bino-like neutralino, the muon sneutrino and the
mostly right-handed smuon. From these pole masses the five SUSY mass
parameters are determined by iteration, such that the on-shell
conditions are satisfied.\footnote{
  For mass spectra where the splitting of the chargino pole masses
  is small compared to the off-diagonal elements in the chargino mass matrix it can
  happen that on-shell values for $M_2$ and $\mu$, which reproduce the
  chargino pole masses at the tree-level, do not exist.  In this case
  \OURPROGRAM\ prints a warning message to \code{stderr} and to
  \code{SPINFO[3]} if \SPHENO\ or NMSSMTools compliant output has been
  selected.  The warning will tell the remaining absolute difference
  (in GeV) between the pole and tree-level chargino masses after the
  iteration has finished.}
The corresponding \DRbar\ values of the
SUSY mass parameters provided by SLHA are ignored.

At the two-loop level, the full spectrum of the MSSM enters, and
$\amu^{\text{SUSY}}$ depends on parameters of all sectors. Here we
highlight the parameters
\begin{align}
M_{Q3}, M_{U3},A^u_{3,3};
M_A
\end{align}
which are of particular phenomenological interest.  Generally
$M_{Qi}$, $M_{Ui}$, $M_{Di}$, $A^u_{i,i}$ denote the left- and
right-handed squark mass parameters and the up-type trilinear coupling
of generation $i$.  Particularly the stops and their masses and
mixings enter via the contributions of Fig.~\ref{fig:diagramsA}(right) and
of Fig.~\ref{fig:diagramsB}(right).  The CP-odd Higgs-Boson mass $M_A$ and
all other heavy Higgs-Boson masses enter via the class 2L(a)
contribution of Fig.~\ref{fig:diagramsA}(right).

The renormalization scheme for those
two-loop parameters is left unspecified, and the parameters are read
directly from the respective input files.

\section{Code details and usage}
\label{sec:usage}

\subsection{Quick start guide}
\label{sec:firstusage}

From the Hepforge page \url{https://gm2calc.hepforge.org} the source
code of \OURPROGRAM\ can be obtained as\footnote{%
  The Hepforge page also features an online calculator
  \url{https://gm2calc.hepforge.org/online.php}, where the
  program can be run without downloading it.}  
\begin{lstlisting}[style=Mybash]
$ wget http://www.hepforge.org/archive/gm2calc/gm2calc-1.0.0.tar.gz
$ tar -xf gm2calc-1.0.0.tar.gz
$ cd gm2calc-1.0.0/
\end{lstlisting}
 To compile the program run
\begin{lstlisting}[style=Mybash]
$ make
\end{lstlisting}
Apart from a C++ compiler, the following headers are required to
compile \OURPROGRAM\@: BOOST (available at
\url{https://www.boost.org}) and Eigen (available at
\url{http://eigen.tuxfamily.org}).  Please refer to the \code{README}
file for customization of the used C++ compiler as well as the
locations of the BOOST and Eigen header files.

\OURPROGRAM\ can be run from the command line by providing a file
containing the input parameters.  Two different input formats are
accepted as explained in Section~\ref{sec:input_parameters}: the first
is the standard SLHA version~1 format \cite{Skands:2003cj}.  Using default settings and the provided
sample SLHA input file \code{input/example.slha}, \OURPROGRAM\ can be
run as
\begin{lstlisting}[style=Mybash]
$ bin/gm2calc.x --slha-input-file=input/example.slha
\end{lstlisting}

This input format is especially useful for cases where e.\,g.\ a spectrum
generator writes an SLHA output file to \code{stdout}, which can then
be streamed into \OURPROGRAM\ using the dash \code{-} as special input-file name.  For
example, using the spectrum generator \SOFTSUSY\ with executable
\code{softpoint.x} from the \SOFTSUSY\ directory with path \code{/path/to}, one can write
\begin{lstlisting}[style=Mybash]
$ /path/to/softpoint.x leshouches < /path/to/inOutFiles/lesHouchesInput | \
    bin/gm2calc.x --slha-input-file=-
\end{lstlisting}
Here \SOFTSUSY\ reads the SLHA input in form of a stream from one of
its example input files \code{inOutFiles/lesHouchesInput}
(CMSSM parameter point 10.1.1~\cite{AbdusSalam:2011fc}).  The output
of \SOFTSUSY\ is then piped into \OURPROGRAM\ which calculates $\amu$.

The second possible input format is the \OURPROGRAM-specific one.  The
example input file \code{input/example.gm2} in this format can be
passed to the program as
\begin{lstlisting}[style=Mybash]
$ bin/gm2calc.x --gm2calc-input-file=input/example.gm2
\end{lstlisting}
In the following we first present the input formats and possible
options in detail and then describe the output.

\subsection{General options}
\label{sec:general-input}

While the first input option is an SLHA file, the GM2Calc-specific input file also has a structure which is similar to the
SLHA standard.  The input files are organized in the form of blocks which
start with the \code{Block} identifier followed by the block name and
an optional scale specification.  The parameters are stored linewise
inside the blocks.  To distinguish the different parameters in a
certain block, each line starts with one or more indices,
followed by the corresponding parameter value.

Common to both input formats are the options for the output format, the
precision of the $\amu$ calculation, and the SM input parameters.  The
output format of \OURPROGRAM\ as well as the precision of the
calculation of $\amu$ can be customized by adding a dedicated
\code{GM2CalcConfig} block to the input file.  The
\code{GM2CalcConfig} block with the default settings has the form
\begin{lstlisting}
Block GM2CalcConfig
     0     3     # output format (0 = minimal, 1 = detailed,
                 #  2 = NMSSMTools, 3 = SPheno, 4 = GM2Calc)
     1     2     # loop order (0, 1 or 2)
     2     1     # disable/enable tan(beta) resummation (0 or 1)
     3     0     # force output (0 or 1)
     4     0     # verbose output (0 or 1)
     5     0     # calculate uncertainty
\end{lstlisting}
The entry \code{GM2CalcConfig[0]} specifies the form of the program
output.\footnote{Here and in the following we refer to a parameter
  value inside a block as \code{BLOCKNAME[INDEX]}, where
  \code{BLOCKNAME} is the name of the block and \code{INDEX} is the
  index referring to a particular line in the block.}  The default
value is \code{3} for the case of SLHA input, but \code{1} for the
case of \OURPROGRAM-specific input. The different output formats are
illustrated in Section \ref{sec:output}.  In \code{GM2CalcConfig[1]}
the loop order of the calculation can be selected (default: \code{2}).
The $\tan\beta$ resummation can be switched on/off by setting the flag
\code{GM2CalcConfig[2]} to \code{1} or \code{0}, respectively
(default: \code{1}).  With the flag \code{GM2CalcConfig[3]}, the
program can be forced to print an output, even if a physical problem
has occurred (for example if tachyons occur in the spectrum)
(default: \code{0}).  Additional information about the internally
performed calculational steps, for example the determination of
on-shell parameters from pole masses,
can be displayed by setting the flag \code{GM2CalcConfig[4]}
to \code{1} (default: \code{0}).
By setting entry \code{GM2CalcConfig[5]} to \code{1} (default:
\code{0}), the theory uncertainty $\delta\amu^{\text{SUSY}}$ is
computed using Eq.~\eqref{error}.  If \code{GM2CalcConfig[0]} has
been set to \code{0} (minimal output), the calculated uncertainty is
written to \code{stdout} as a single number, instead of the value of
$\amu^{\text{SUSY}}$.  If \code{GM2CalcConfig[0]} has been set to
\code{2}, \code{3} or \code{4} (SLHA-compliant output formats),
$\delta\amu^{\text{SUSY}}$ is written to \code{GM2CalcOutput[1]}.

To specify the Standard Model input parameters the \code{SMINPUTS} block
must be given, as defined in Ref.~\cite{Skands:2003cj}.  In addition,
the $W$ pole mass can be given in \code{SMINPUTS[9]}.  Using
\OURPROGRAM's default values for the Standard Model input parameters,
the \code{SMINPUTS} block reads
\begin{lstlisting}
Block SMINPUTS
     3     0.1184           # alpha_s(MZ) SM MS-bar [relevant at 2L]
     4     91.1876          # M_Z(pole)             [relevant at 1L]
     5     4.18             # m_b(m_b) SM MS-bar    [relevant at 2L]
     6     173.34           # M_top(pole)           [relevant at 2L]
     7     1.777            # M_tau(pole)           [relevant at 2L]
     9     80.385           # M_W(pole)             [relevant at 1L]
    13     0.1056583715     # M_muon(pole)          [relevant at 1L]
\end{lstlisting}
Other SM input parameters are not needed and therefore ignored if
specified. In particular, the Standard Model
\MSbar\ value of the inverse fine-structure constant,
$[\alpha_\text{em}^{\MSbar}(M_Z)]^{-1}$, usually given in
\code{SMINPUTS[1]}, is ignored.

The default values of $\alpha(M_Z)$ and $\alpha(0)$ as defined in
Section \ref{sec:input_parameters} can be overridden by providing new
values in the \code{GM2CalcInput} block.  With the
default values this block reads
\begin{lstlisting}
Block GM2CalcInput
     1     0.00775531       # alpha(MZ)
     2     0.00729735       # alpha(0)     [used in 2L photonic contributions]
\end{lstlisting}
%

\subsection{Usage with SLHA input format}
\label{sec:usage-slha}

\OURPROGRAM\ can be run by providing the input parameters in
SLHA-compliant format~\cite{Skands:2003cj}.  An example SLHA input
file suitable for \OURPROGRAM\ can be found in
\ref{sec:example-slha-file} (MSSM parameter point in the style of
BM1~\cite{Fargnoli:2013zda}). This file also contains information
about the loop order at which the different entries become relevant.
Some parameter values, marked as irrelevant, are only included for the
sake of completeness and not required by \OURPROGRAM.

The entries are as follows: the running MSSM \DRbar\
parameters must be given in the SLHA parameter blocks \code{HMIX},
\code{AU}, \code{AD}, \code{AE} and \code{MSOFT} as defined
in Ref.~\cite{Skands:2003cj}.  The required pole masses of SUSY particles must
be provided in the \code{MASS} block as defined again
in Ref.~\cite{Skands:2003cj}.
In particular, the two smuon pole masses and the muon sneutrino pole
mass must be given in \code{MASS[1000013]}, \code{MASS[2000013]} and
\code{MASS[1000014]}, respectively, as defined in SLHA version~1
\cite{Skands:2003cj}.  Inter-generation sfermion mixing, as defined in
SLHA version~2 \cite{Allanach:2008qq}, would lead to a re-assignment
of the sfermion pole masses to the entries of the \code{MASS} block,
which is currently not supported.
Additionally, the $W$ pole mass can be given in \code{MASS[24]}, which
overrides the value provided in \code{SMINPUTS[9]}.

We strongly recommend to provide the \DRbar\ values of $\mu$, $M_1$
and $M_2$ in the SLHA input file.  They will be treated as an initial
guess for the corresponding on-shell values, which are determined
iteratively from the two chargino and the bino-like neutralino pole
masses as described in Section \ref{sec:input_parameters}.  If the
\DRbar\ values of $\mu$, $M_1$ and $M_2$ are omitted, they will be
treated as being zero, which might result in a bad initial guess for
the corresponding on-shell values.


\subsection{Usage with GM2Calc-specific input format}
\label{sec:usage-onshell}

As an alternative to the SLHA input format, \OURPROGRAM\ can be run by
directly providing the input parameters in the renormalization scheme
presented in Section~\ref{sec:input_parameters}.
Using this \OURPROGRAM-specific input format, the MSSM parameters
must be provided in a
dedicated \code{GM2CalcInput} block, which is exemplified by the complete
input file in \ref{sec:example-gm2calc-file}.


\subsection{Output formats}
\label{sec:output}

The output format of the program can be selected by setting the
variable \code{GM2CalcConfig[0]}. In case of SLHA input,
\code{GM2CalcConfig[0]} is by default set to \code{3} (\SPHENO\
output). Then the complete SLHA input is written to \code{stdout} with
the calculated value of $\amu$ added to
\code{SPhenoLowEnergy[21]}:
\begin{lstlisting}[style=Mybash]
$ bin/gm2calc.x --slha-input-file=input/example.slha
\end{lstlisting}
\begin{lstlisting}
...
Block SPhenoLowEnergy
    21     2.30368508E-09   # Delta(g-2)_muon/2
\end{lstlisting}
The dots  abbreviate the SLHA input. 
If the \code{SPhenoLowEnergy} block or the entry \code{SPhenoLowEnergy[21]} do not exist, they are
created and appended to the output.  If the entry \code{SPhenoLowEnergy[21]}
already exists, it is overwritten by the value of $\amu$ calculated by
\OURPROGRAM\@.

If \code{GM2CalcConfig[0]} is set to \code{2} (NMSSMTools output), the
behaviour is almost the same, except for $a_\mu$ not being written to
\code{SPhenoLowEnergy[21]} but to \code{LOWEN[6]}.\footnote{Note that this
  option only changes the output {\em format}. The resulting value
  still corresponds to $\amu$ in the MSSM and not the NMSSM.}
  Providing the SLHA input of the example given in
\ref{sec:example-slha-file}, the NMSSMTools output reads:\footnote{%
  To make the example self-contained, the piping functionality of the
  Bourne shell is used to add the \code{GM2CalcConfig[0]} setting to the SLHA
  input file.}
\begin{lstlisting}[style=Mybash]
$ { cat input/example.slha;
    cat <<EOF
Block GM2CalcConfig
     0     2     # NMSSMTools output
EOF
  } | bin/gm2calc.x --slha-input-file=-
\end{lstlisting}
\begin{lstlisting}
...
Block LOWEN
     6     2.30368508E-09   # Delta(g-2)_muon/2
\end{lstlisting}
The dots again abbreviate the SLHA input which is written to the
output.
By setting \code{GM2CalcConfig[0]} to \code{4} the output is the same
as above, except that the value of $\amu$ is written to
\code{GM2CalcOutput[0]}.  This choice is useful if interference with
\SPHENO\ and NMSSMTools must be avoided.

If \code{GM2CalcConfig[0]} is set to \code{0} (minimal output), the
program writes only the value of $\amu$ to \code{stdout}.  For the
SLHA input given in \ref{sec:example-slha-file} the minimal output
looks as follows:
\begin{lstlisting}[style=Mybash]
$ { cat input/example.slha;
    cat <<EOF
Block GM2CalcConfig
     0     0     # minimal output
EOF
  } | bin/gm2calc.x --slha-input-file=-
\end{lstlisting}
\begin{lstlisting}
2.30368508e-09
\end{lstlisting}

If \code{GM2CalcConfig[0]} is set to \code{1} (detailed output),
\OURPROGRAM\ writes detailed information about the different
contributions to $\amu$ to \code{stdout}:
\\
\begin{minipage}{\linewidth}
\begin{lstlisting}[style=Mybash]
$ { cat input/example.slha;
    cat <<EOF
Block GM2CalcConfig
     0     1     # detailed output
EOF
  } | bin/gm2calc.x --slha-input-file=-
\end{lstlisting}
\end{minipage}
\begin{lstlisting}
====================================================================
   amu (1-loop + 2-loop best) =  2.30368508e-09 +- 2.33327662e-10
====================================================================

==============================
   amu (1-loop) corrections
==============================

full 1L with tan(beta) resummation:
   chi^0     -2.41810081e-10
   chi^+-     2.66183984e-09
   -------------------------------
   sum        2.42002976e-09 (105.1% of full 1L + 2L result)

full 1L without tan(beta) resummation:
              2.24788956e-09

1L approximation with tan(beta) resummation:
   W-H-nu     2.69541309e-09
   W-H-muL   -4.11041944e-10
   B-H-muL    1.04874082e-10
   B-H-muR   -2.26475517e-10
   B-muL-muR  2.78877323e-10
   -------------------------------
   sum        2.44164703e-09

==============================
   amu (2-loop) corrections
==============================

2L best with tan(beta) resummation:
             -1.16344676e-10 (-5.1% of full 1L + 2L result)

2L best without tan(beta) resummation:
             -1.05961600e-10

photonic with tan(beta) resummation:
   chi^0      1.98555488e-11
   chi^+-    -2.16650643e-10
   -------------------------------
   sum       -1.96795094e-10 (-8.5% of full 1L + 2L result)

fermion/sfermion approximation with tan(beta) resummation:
   W-H-nu     7.32826955e-11
   W-H-muL   -1.11753785e-11
   B-H-muL   -1.27259647e-12
   B-H-muR   -2.91371660e-12
   B-muL-muR  1.14372064e-11
   -------------------------------
   sum        6.93582103e-11 (3.0% of full 1L + 2L result)

2L(a) (1L insertions into 1L SM diagram) with tan(beta) resummation:
   sfermion   1.11553353e-16
   cha^+-     1.10920961e-11
   -------------------------------
   sum        1.10922077e-11 (0.5% of full 1L + 2L result)

tan(beta) correction:
   amu(1L) * (1 / (1 + Delta_mu) - 1) =  1.71751841e-10 (7.6%)
\end{lstlisting}
The detailed output format is used by default if the input is provided
in the \OURPROGRAM-specific format and \code{GM2CalcConfig[0]} has not
been set.

\section{Examples of how to use \OURPROGRAM}
\label{sec:examples}

In this section several practical applications of \OURPROGRAM\ are
shown with the help of a few examples.

\subsection{Using input from spectrum generator, piping output
  to external programs}

If a spectrum generator writes an SLHA output to \code{stdout},
this output can be streamed into \OURPROGRAM\ using the dash \code{-}
as special input-file
name.  As mentioned in Section \ref{sec:firstusage}, when
using the spectrum generator \SOFTSUSY\ one can write
\begin{lstlisting}[style=Mybash]
$ /path/to/softpoint.x leshouches < /path/to/inOutFiles/lesHouchesInput | \
    bin/gm2calc.x --slha-input-file=-
\end{lstlisting}
\begin{lstlisting}
...
Block SPhenoLowEnergy
    21     8.31313424E-10   # Delta(g-2)_muon/2
\end{lstlisting}
Here \code{softpoint.x} is the \SOFTSUSY\ executable which reads the
SLHA input in form of a stream from one of \SOFTSUSY's default input
files \code{inOutFiles/lesHouchesInput} (CMSSM parameter point
10.1.1~\cite{AbdusSalam:2011fc}).  The output of \SOFTSUSY\ is then
piped into \OURPROGRAM\ which calculates $\amu$ and writes the result
to \code{SPhenoLowEnergy[21]}.

In the example given above the default settings of \OURPROGRAM\ are
used, since the output of \SOFTSUSY\ does not contain any
\OURPROGRAM-specific blocks such as \code{GM2CalcConfig} or
\code{GM2CalcInput}. If an additional input block like
\code{GM2CalcConfig} shall be passed to \OURPROGRAM, a simple and
self-contained way is to modify the command to
\begin{lstlisting}[style=Mybash]
$ { /path/to/softpoint.x leshouches < /path/to/inOutFiles/lesHouchesInput;
    cat <<EOF
Block GM2CalcConfig
     0     0     # minimal output
EOF
  } | bin/gm2calc.x --slha-input-file=-
\end{lstlisting}
\begin{lstlisting}
8.31313424e-10
\end{lstlisting}

\begin{figure}[htb]
  \centering
  \includegraphics[width=.49\linewidth]{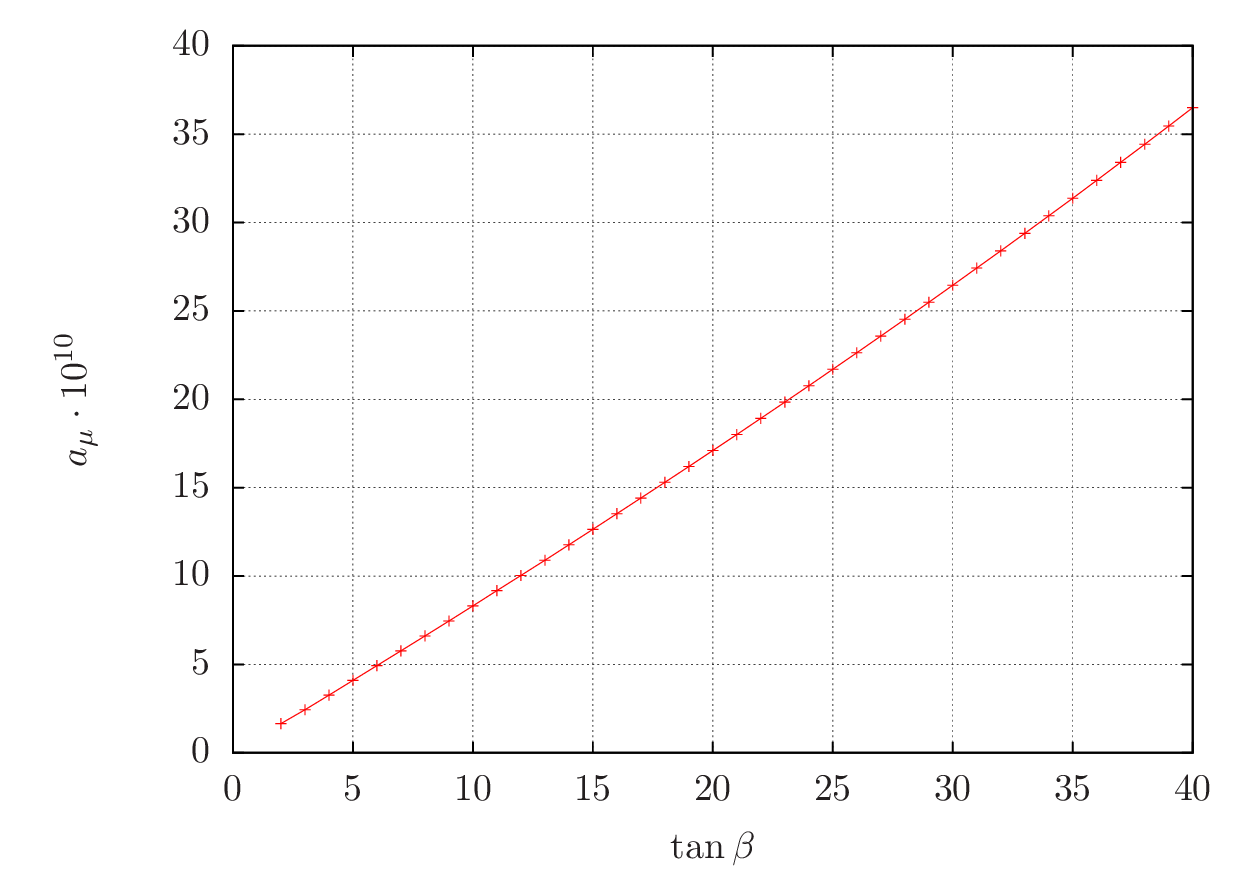}
  \hfill
  \includegraphics[width=.49\linewidth]{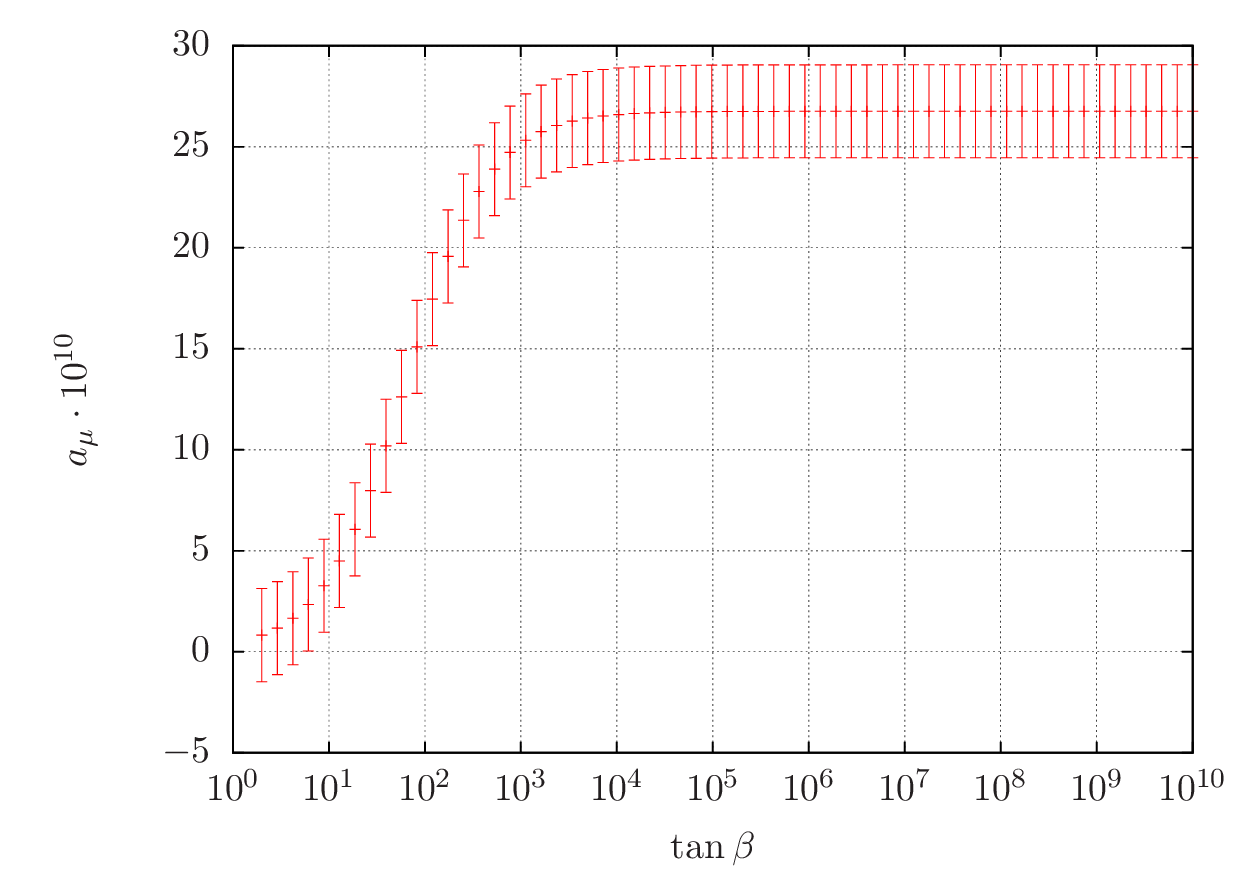}
  \caption{The two graphics show $\amu^{\text{SUSY}}$ as a function of
    $\tan\beta$ for the MSSM calculated according to Eq.\
    \eqref{amuSUSYdecomposition}.  In the left plot,
    $\amu^{\text{SUSY}}$ is calculated for the parameter point
    10.1.1~\cite{AbdusSalam:2011fc}.  The right plot shows
    $\amu^{\text{SUSY}}$ together with the uncertainty Eq.\
    \eqref{error} for benchmark point 1 from Ref.\
    \cite{Bach:2015doa}, where the on-shell parameters relevant at the
    one-loop level are $\mu = -M_2 = 30 \unit{TeV}$, $M_1 = M_{L2} =
    M_{E2} = 1 \unit{TeV}$, $A^e_{2,2}(Q) = 0$ and the parameters
    relevant at the two-loop level are set to $M_3 = M_A = M_{Q1,Q2} =
    M_{U1,U2} = M_{D1,D2} = M_{L1} = M_{E1} = 1 \unit{TeV}$, $M_{Q3} =
    M_{U3} = M_{D3} = M_{L3} = M_{E3} = 3 \unit{TeV}$, $A^f_{ij} = 0$
    and $Q=866.36\unit{GeV}$.  Due to the $\tan\beta$ resummation,
    $\amu^{\text{SUSY}}$ approaches the finite maximum value $\amu^{\text{SUSY}} = 26.8\cdot 10^{-10}$ in the
    limit $\tan\beta\rightarrow\infty$.
    This finite limit of $\amu^{\text{SUSY}}$ decomposes into the
    individual one- and two-loop contributions
    $[\amuSUOL]_{t_\beta\text{-resummed}} = 28.1 \cdot 10^{-10}$,
    $[\amuphotonic]_{t_\beta\text{-resummed}} = -2.3 \cdot 10^{-10}$,
    $[\amuFSf]_{t_\beta\text{-resummed}} = 0.9 \cdot 10^{-10}$,
    $[\amuTLa]_{t_\beta\text{-resummed}} < 10^{-13}$.}
  \label{fig:amu-tb}
\end{figure}

By providing extra input blocks and writing
loops at the command line, one can easily perform
parameter scans at the command line without the need for creating
temporary files.  The output of a scan can directly be piped to a
program for visualization, e.\,g.\ gnuplot. For instance, the following script first
defines four auxiliary functions, each of which can easily be extended
for further use. The script then calls the functions in a loop over
$\tan\beta$ and produces a plot similar to
\figurename\ \ref{fig:amu-tb}(left).
It is easy to modify the script for more sophisticated scans like the
one shown in \figurename\ \ref{fig:amu-tb}(right).
\begin{lstlisting}[style=Mybash]
softpoint_with_block() {
  { cat -; printf "Block $1\n\t$2\t$3\n"; } | /path/to/softpoint.x leshouches
}

amu_from_slha() {
  { cat -; printf "Block GM2CalcConfig\n\t0\t0\n"; } | # minimal output
    bin/gm2calc.x --slha-input-file=-
}

amu_with_block() {
  softpoint_with_block "$@" | amu_from_slha
}

amu_for_tanbeta() {
  amu_with_block < /path/to/inOutFiles/lesHouchesInput MINPAR 3 $1 # tan(beta) at MZ
}

{ echo "set xlabel \"tan(beta)\";
        set ylabel \"amu\";
        plot '-' u 1:2 w linespoints t ''"
  for tb in $(seq 2 40); do
    echo "$tb $(amu_for_tanbeta $tb)"
  done
} | gnuplot -p
\end{lstlisting}

\subsection{C++ interface}

\OURPROGRAM\ provides a C++ interface which allows users to embed the
calculation of $\amu$ into an existing code or create a custom C++
program that calculates $\amu$.  The following source-code listing
shows an example of a C++ program which
calculates $\amu$ up to the two-loop level including $\tan\beta$
resummation using the \OURPROGRAM-specific input format.
\begin{lstlisting}[language=C++, caption=example-gm2calc.cpp]
#include "gm2_1loop.hpp"
#include "gm2_2loop.hpp"
#include "MSSMNoFV_onshell.hpp"
#include <iostream>

gm2calc::MSSMNoFV_onshell setup() {
   gm2calc::MSSMNoFV_onshell model;

   const Eigen::Matrix<double,3,3> UnitMatrix
      = Eigen::Matrix<double,3,3>::Identity();

   // fill DR-bar parameters
   model.set_TB(10);                      // 1L
   model.set_Ae(1,1,0);                   // 1L

   // fill on-shell parameters
   model.set_Mu(350);                     // 1L
   model.set_MassB(150);                  // 1L
   model.set_MassWB(300);                 // 1L
   model.set_MassG(1000);                 // 2L
   model.set_mq2(500 * 500 * UnitMatrix); // 2L
   model.set_ml2(500 * 500 * UnitMatrix); // 1L(smuon)/2L
   model.set_md2(500 * 500 * UnitMatrix); // 2L
   model.set_mu2(500 * 500 * UnitMatrix); // 2L
   model.set_me2(500 * 500 * UnitMatrix); // 1L(smuon)/2L
   model.set_Au(2,2,0);                   // 2L
   model.set_Ad(2,2,0);                   // 2L
   model.set_Ae(2,2,0);                   // 2L
   model.set_MA0(1500);                   // 2L
   model.set_scale(454.7);                // 2L

   // calculate mass spectrum
   model.calculate_masses();

   return model;
}

int main() {
   gm2calc::MSSMNoFV_onshell model(setup());

   const double amu =
      + gm2calc::calculate_amu_1loop(model)
      + gm2calc::calculate_amu_2loop(model);

   std::cout << "amu = " << amu << std::endl;

   return 0;
}
\end{lstlisting}
The object \code{model} contains the model parameters. The \code{setup()} function
initializes this object by first defining input
parameters in the \OURPROGRAM-specific input format and then
calculating the tree-level mass spectrum with the provided function
\code{calculate_masses()}. Afterwards, the \code{main()} function uses the
initialized object to calculate $\amu$. 

The listed source code can be compiled using a C++ compiler and
linking the static library \code{libgm2calc.a} as follows:
\begin{lstlisting}[style=Mybash]
$ g++ -Isrc example-gm2calc.cpp src/libgm2calc.a
\end{lstlisting}
Afterwards, the created executable can be run via
\begin{lstlisting}[style=Mybash]
$ ./a.out
\end{lstlisting}
which will produce the output:
\begin{lstlisting}
amu = 7.96432e-10
\end{lstlisting}

It is also possible to use the
SLHA input format at the C++ level, which is exemplified by the
following source-code listing.
Like in the previous example, there is a \code{model} object and a
\code{setup()} function. But now, the
\code{setup()} function fills the pole masses
of the relevant SUSY particles and \DRbar\ Lagrangian parameters into
the \code{model} object.  Afterwards, the provided function
\code{convert_to_onshell()} is called, which determines the on-shell
parameters $\mu$, $M_1$, $M_2$, $M_{E2}$ and $M_{L2}$ from the corresponding
pole masses as described in Section~\ref{sec:input_parameters}.
Internally, this function
finally calculates the tree-level mass spectrum
using these on-shell parameters.  In the \code{main()} function, this
tree-level mass spectrum is used to calculate $\amu$ up to the
two-loop level, including $\tan\beta$ resummation, and the result is
printed to \code{stdout}.
\begin{lstlisting}[language=C++, caption=example-slha.cpp]
#include "gm2_1loop.hpp"
#include "gm2_2loop.hpp"
#include "MSSMNoFV_onshell.hpp"
#include <iostream>

gm2calc::MSSMNoFV_onshell setup() {
   gm2calc::MSSMNoFV_onshell model;

   const Eigen::Matrix<double,3,3> UnitMatrix
      = Eigen::Matrix<double,3,3>::Identity();

   // fill pole masses
   model.get_physical().MSvmL   =  5.18860573e+02; // 1L
   model.get_physical().MSm(0)  =  5.05095249e+02; // 1L
   model.get_physical().MSm(1)  =  5.25187016e+02; // 1L
   model.get_physical().MChi(0) =  2.01611468e+02; // 1L
   model.get_physical().MChi(1) =  4.10040273e+02; // 1L
   model.get_physical().MChi(2) = -5.16529941e+02; // 1L
   model.get_physical().MChi(3) =  5.45628749e+02; // 1L
   model.get_physical().MCha(0) =  4.09989890e+02; // 1L
   model.get_physical().MCha(1) =  5.46057190e+02; // 1L
   model.get_physical().MAh(1)  =  1.50000000e+03; // 2L

   // fill DR-bar parameters
   model.set_TB(40);                        // 1L
   model.set_Mu(500);                       // initial guess
   model.set_MassB(200);                    // initial guess
   model.set_MassWB(400);                   // initial guess
   model.set_MassG(2000);                   // 2L
   model.set_mq2(7000 * 7000 * UnitMatrix); // 2L
   model.set_ml2(0, 0, 500 * 500);          // 2L
   model.set_ml2(1, 1, 500 * 500);          // irrelevant
   model.set_ml2(2, 2, 500 * 500);          // 2L
   model.set_md2(7000 * 7000 * UnitMatrix); // 2L
   model.set_mu2(7000 * 7000 * UnitMatrix); // 2L
   model.set_me2(0, 0, 500 * 500);          // 2L
   model.set_me2(1, 1, 500 * 500);          // initial guess
   model.set_me2(2, 2, 500 * 500);          // 2L
   model.set_Au(2, 2, 0);                   // 2L
   model.set_Ad(2, 2, 0);                   // 2L
   model.set_Ae(1, 1, 0);                   // 1L
   model.set_Ae(2, 2, 0);                   // 2L
   model.set_scale(1000);                   // 2L

   // convert DR-bar parameters to on-shell
   model.convert_to_onshell();

   return model;
}

int main() {
   gm2calc::MSSMNoFV_onshell model(setup());

   const double amu =
      + gm2calc::calculate_amu_1loop(model)
      + gm2calc::calculate_amu_2loop(model);

   std::cout << "amu = " << amu << std::endl;

   return 0;
}
\end{lstlisting}
This program will produce the output:
\begin{lstlisting}
amu = 2.33925e-09
\end{lstlisting}

\section{Summary and final comments}
\label{sec:summary}

We have presented \OURPROGRAM, a C++ program to calculate the
anomalous magnetic moment of the muon $\amu$ in the MSSM\@. It includes
one-loop and Barr-Zee-like two-loop contributions as well as more recently
computed two-loop  photonic and fermion/sfermion-loop 
corrections. By default, $\tan\beta$ resummation is performed, allowing for
arbitrarily high values of $\tan\beta$.
The program input can be provided in
either an SLHA (version~1) compliant or a \OURPROGRAM-specific format.
Internally, \OURPROGRAM\ uses a physical, on-shell renormalization
scheme to minimize two-loop contributions and theory uncertainties.

We have given simple usage examples where \OURPROGRAM\ is run on its
own or using the SLHA output of a spectrum generator, and we have
illustrated how the output of \OURPROGRAM\ can be passed to
external programs like gnuplot. We have also discussed sample C++ programs which use the
\OURPROGRAM\ libraries and routines. All examples and their
explanations can also be found on the web site
\url{https://gm2calc.hepforge.org}. This web site further provides an
online calculator, which allows users to type in parameters and compute
$\amu^{\text{SUSY}}$ interactively without downloading the code.

The program has been thoroughly validated against the original
routines of
Refs.~\cite{HSW03,HSW04,DSreview,vonWeitershausen:2010zr,Fargnoli:2013zda,Fargnoli:2013zia,Marchetti:2008hw,Bach:2015doa}. 
The estimate of the theory uncertainty given in
Section~\ref{sec:error} follows the analysis of
Ref.~\cite{DSreview} of the missing contributions and is specific to
the computation in the chosen renormalization scheme. When comparing
\OURPROGRAM\ to other codes/evaluations, differences can arise for
several reasons. Particularly, \OURPROGRAM\ differs from evaluations
which use the
\DRbar\ scheme to define the masses entering
the one-loop contributions. In the latter case  $\amuSUOL$ depends on the 
\DRbar\ scale and there are potentially very large
two-loop corrections discussed in Section~\ref{sec:contributions}.
Clearly, a more trivial reason for numerical differences is a different choice of
the implemented two-loop contributions. To our knowledge the photonic, the
fermion/sfermion-loop and the $\tan\beta$-resummation corrections are
implemented in this form for the first time. Each of them can amount to
${\cal O}(10\%)$ corrections or more in parts of the parameter space.

\section*{Acknowledgments}
We acknowledge financial support by the German Research Foundation DFG
through Grants STO876/2-1 and STO876/6-1 and
by the Collaborative Research Center SFB676 of the DFG, ``Particles,
Strings and the early Universe''. The work of P.A.~was supported by the ARC Centre of Excellence for Particle Physics at the Terascale.
J.P. acknowledges support from the MEC and FEDER (EC) Grant
FPA2011--23596 and the Generalitat Valenciana under Grant PROMETEOII/2013/017.
A.V.\ would like to thank
Bj\"orn Sarrazin for testing the alpha version of \OURPROGRAM.

\begin{appendix}
\section{Detailed contributions}
The one-loop and the photonic two-loop contributions can be written as
\begin{subequations}
\begin{align}
\amuSUOL&=a_\mu^{\chi^\pm}+a_\mu^{\chi^0} ,
\\
\amuphotonic &=
a^{{\chi^\pm\,(\gamma)}}_{\mu} 
+
a^{{\chi^0\,(\gamma)}}_{\mu} ,
\end{align}
\end{subequations}
and they are  implemented in the form given in
Refs.~\cite{vonWeitershausen:2010zr,Fargnoli:2013zda,Fargnoli:2013zia}:
\newcommand{\snumu}{{\tilde \nu}_{\mu}}
\newcommand{\smum}{{\tilde \mu}_{m}}
\begin{subequations}
\begin{align}
a_\mu^{\chi^\pm}&=\sum_{i}
 \frac{1}{16 \pi^2}\,\frac{m_{\mu}^{2}}{m_{\snumu}^{2}}\Big\{
\frac{1}{12}{\cal A}^{c\,+}_{ii \snumu}
F_1^C(x)
+\frac{m_{\chi^-_i}}{3m_{\mu}}
{\cal B}^{c\,+}_{ii \snumu} F_2^C(x)
\Big\}
,\\
\label{eq:oneloopneu} 
a_\mu^{\chi^0}&=\sum_{i,m}
\frac{-1}{16\pi^2}\frac{m_{\mu}^{2}}{m_{\smum}^2}
\Big\{
\frac{1}{12}{\cal A}^{n\,+}_{ii \smum}
F_1^N(x)
+\frac{m_{\chi^0_i}}{6m_{\mu}}
{\cal B}^{n\,+}_{ii \smum} F_2^N(x)
\Big\} , 
\end{align}
\begin{align}
a^{{\chi^\pm\,(\gamma)}}_{\mu} = \sum_{i} \frac{1}{16\pi^2} \frac{\alpha(0)}{4\pi}
\frac{m_\mu^2}{m_{\tilde{\nu}_\mu}^2}
\Bigg[
  & \bigg(
    \frac{1}{12}{\cal A}_{ii\snumu}^{c\,+} F_1^C(x) 
    + \frac{2 m_{\chi^-_i}}{3 }  \frac{{\cal B}_{ii\snumu}^{c\,+}}{2m_\mu} F_2^C(x)
  \bigg) 16\log \frac{m_\mu}{m_{\tilde\nu_\mu}} 
\nonumber\\
  - &\bigg(
    \frac{47 }{72} {\cal A}_{ii\snumu}^{c\,+} F_{3}^C(x) 
    + \frac{122 m_{\chi^-_i}}{9} \frac{{\cal B}_{ii\snumu}^{c\,+}}{2m_\mu} F_{4}^C(x)
  \bigg) 
\nonumber\\
  - & \bigg(
    \frac{1}{2} {\cal A}_{ii\snumu}^{c\,+} F_1^C(x)
    + { 2m_{\chi^-_i}} \frac{{\cal B}_{ii\snumu}^{c\,+}}{2m_\mu} F_2^C(x)
  \bigg) {\rm L}(m_{\tilde\nu_\mu}^2) \Bigg],
\label{eq:chares}
 \\
a^{{\chi^0\,(\gamma)}}_{\mu} = \sum_{i,m}\frac{1}{16\pi^2} \frac{\alpha(0)}{4\pi} 
\frac{m_\mu^2}{m_{\tilde{\mu}_m}^2}
\Bigg[
  &\bigg(
    -\frac{1}{12} {\cal A}_{ii\smum}^{n\,+} F_1^N(x)
    +\frac{m_{\chi^0_i}}{3}  \frac{-{\cal B}_{ii\smum}^{n\,+}}{2m_\mu} F_2^N(x)
  \bigg) 16\log \frac{m_\mu}{m_{\tilde\mu_m}} 
\nonumber\\
  - &\bigg(
    - \frac{35 }{72}  {\cal A}_{ii\smum}^{n\,+} F_{3}^N(x)
    + \frac{16 m_{\chi^0_i}}{9} \frac{-{\cal B}_{ii\smum}^{n\,+}}{2m_\mu} F_{4}^N(x)
  \bigg) 
\nonumber\\
  + &\bigg(
    \frac{1}{4 } {\cal A}_{ii\smum}^{n\,+} F_1^N(x)
  \bigg) {\rm L}(m_{\tilde\mu_m}^2) \Bigg] ,
\label{eq:neures}
\end{align}
\end{subequations}
where
\newcommand{\acplusminus}{
 z^{L}_{i{\tilde f}_{k}} \, z^{L \, *}_{j{\tilde f}_{k}} \pm
 z^{R}_{i{\tilde f}_{k}} \, z^{R \, *}_{j{\tilde f}_{k}}
} 
\newcommand{\bcplusminus}{
 z^{L}_{i{\tilde f}_{k}} \, z^{R \, *}_{j{\tilde f}_{k}} \pm
 z^{R}_{i{\tilde f}_{k}} \, z^{L \, *}_{j{\tilde f}_{k}}
}
\begin{subequations}
\begin{align}
{\cal A}^{z \, \pm}_{i j {\tilde f}_{k}} &\equiv \acplusminus, \\ 
{\cal B}^{z \, \pm}_{i j {\tilde f}_{k}} &\equiv \bcplusminus,
\end{align}
\end{subequations}
with~$z\in\{c,n\}$. All conventions have been unified to the ones of
Ref.~\cite{Fargnoli:2013zia}, and the actual couplings $c^{L,R}$ and $n^{L,R}$ of
the charginos/neutralinos to a general lepton $l$ are given by 
\begin{subequations}
\begin{align}
c^{L}_{i\tilde{\nu}_{l}} & = -g_{2}V^{*}_{i1}, \phantom{\frac{1}{1}} \\
c^{R}_{i\tilde{\nu}_{l}} & = y_{l}U_{i2}, \phantom{\frac{1}{1}} \\
n^{L}_{i\tilde{l}_{k}} & = \frac{1}{\sqrt{2}}\left(g_{1}N^{*}_{i1}+g_{2}N^{*}_{i2}\right)U^{\tilde{l}_{}}_{k1}-y_{l}N^{*}_{i3}U^{\tilde{l}_{}}_{k2},\\
n^{R}_{i\tilde{l}_{k}} & = -\sqrt{2}g_{1}N^{}_{i1}U^{\tilde{l}_{}}_{k2}-y_{l}N^{}_{i3}U^{\tilde{l}_{}}_{k1},
\end{align}
\end{subequations}
where $g_{2,1}$ are the SU(2)$\times$U(1) gauge couplings defined via
$\alpha(M_Z)$ and the weak mixing angle
$\sin^2\theta_W=1-M_W^2/M_Z^2$, and $y_l$ denotes the lepton Yukawa
coupling, see Eq.~\eqref{ymuresummation}. The chargino, neutralino,
and slepton mixing matrices $U$, $V$, $N$, $U^{\tilde{l}}$ are defined in
the usual way, see e.\,g.\ Ref.~\cite{DSreview}.
The
arguments of the loop functions are given by $x =
m_{\chi^0_i}^2 / m_{\tilde{\mu}_m}^2$ or $x =
m_{\chi^-_i}^2 / m_{\tilde{\nu}_\mu}^2$, as appropriate.
The one-loop functions
$F_{1,2}^j(x)$  are given in all aforementioned references, the
two-loop functions $F_{3,4}^j(x)$ are given in
Ref.~\cite{vonWeitershausen:2010zr}; they read
\begin{subequations}
\begin{align}
F_1^C(x) &= \frac{ 2 }{(1-x)^4 }\big[
2+3x-6x^2+x^3+6x\log x
\big],\\
F_2^C(x) &= \frac{ 3 }{2(1-x)^3 }\big[
-3+4x-x^2-2\log x
\big],\\
F_1^N(x) &= \frac{ 2 }{(1-x)^4 }\big[
1-6x+3x^2+2x^3-6x^2\log x
\big],\\
F_2^N(x) &= \frac{ 3 }{(1-x)^3 }\big[
1-x^2+2x\log x
\big],
\end{align}
\begin{align}
  F_3^C(x) =\frac{4}{141 (1-x)^4} \Big[
&\big(1-x\big)\big(151 x^2-335 x + 592\big) 
\nonumber \\
   {}+{} & 6 \big(21 x^3- 108x^2-93 x+50\big) \log x \nonumber\\
   {}-{} & 54 x \big(x^2-2 x-2\big) \log^2 x \nonumber\\ 
   {}-{} & 108 x \big(x^2-2 x+12\big) {\rm Li}_2(1-x) \Big], \\
  F_4^C(x) = \frac{-9}{122 (1-x)^3} \Big[
&  8 \big(x^2-3 x+2\big) + \big(11 x^2 - 40x + 5\big) \log x\nonumber  \\
  {}-{} & 2 \big(x^2-2 x-2\big) \log^2 x \nonumber\\
  {}-{} & 4 \big(x^2-2 x+9\big) {\rm Li}_2(1-x) \Big],
\\
F_3^N(x) = \frac{4}{105 (1-x)^4} \Big[
&  \big(1-x\big)\big(-97x^2 -529x +2\big) + 6x^2\big( 13x + 81 \big) \log x \nonumber \\
  {}+{} & 108 x \big(7 x+4\big) {\rm Li}_2(1-x) \Big], \\
F_4^N(x) =\ \frac{-9}{4 (1-x)^3} \ \ \Big[
&  \big(x+3\big) \big(x \log x + x-1\big)+\big(6 x+2\big) {\rm
    Li}_2(1-x)\Big],
\end{align}
and additionally
\begin{align}
{\rm L}(m^2) &= \log\frac{m^2}{Q^2} .
\end{align}
\end{subequations}

The resummation of $\tan\beta$-enhanced contributions is taken into
account by replacing the Yukawa coupling appearing in the coupling
constants by
Eq.~\eqref{ymuresummation}~\cite{Marchetti:2008hw,Bach:2015doa}. The
occurring quantity $\Delta_\mu$ is given by~\cite{Marchetti:2008hw}
\begin{align}
\Delta_\mu =
&-{\mu \ \tan\beta\ }
\frac{g_2^2\ M_2}{16\pi^2}\ I(m_1,m_2,m_{{\tilde{\nu}_\mu}}^{(\infty)})
-{\mu \ \tan\beta\ }
\frac{g_2^2\ M_2}{16\pi^2}\ \frac{1}{2} I(m_1,m_2,m_{\tilde{\mu}_L}^{(\infty)})
\nonumber\\
& -{\mu \ \tan\beta\ } \frac{g_1^2\ M_1}{16\pi^2}\
\Big[
I(\mu,M_1,m_{\tilde{\mu}_R}^{(\infty)})
 -\frac{1}{2}
I(\mu,M_1,m_{\tilde{\mu}_L}^{(\infty)})
-I(M_1,m_{\tilde{\mu}_L}^{(\infty)},m_{\tilde{\mu}_R}^{(\infty)})\Big]
\label{ds-DeltamuResult},
\end{align}
where
\begin{subequations}
\begin{align}
m_{1,2}^2 &=\frac{1}{2}\Big[
(M_2^2 + \mu^2 + 2M_W^2)
\mp\sqrt{(M_2^2 + \mu^2 + 2M_W^2)^2-4M_2^2\mu^2}\Big],  \\
m_{\tilde{\nu}_\mu}^{(\infty)}{}^2& 
=M_{L2}^2- \frac{M_Z^2}{2},
\\
m_{\tilde{\mu}_L}^{(\infty)}{}^2 &= M_{L2}^2-M_Z^2 (s_W^2-\frac12),
 \\
m^{(\infty)}_{\tilde{\mu}_R}{}^2&=M_{E2}^2+M_Z^2 s_W^2,
\\
 I(a,b,c)&=
\frac{
  a^2b^2\log{\frac{a^2}{b^2}}
        +b^2c^2\log{\frac{b^2}{c^2}}
        +c^2a^2\log{\frac{c^2}{a^2}}}{(a^2-b^2)(b^2-c^2)(a^2-c^2)}
.
\end{align}
\end{subequations}
For the Yukawa couplings of the
 $\tau$-lepton and the $b$-quark, $y_\tau$ and $y_b$ respectively, which appear at the two-loop level,
we have implemented the analogue of
Eqs.~(\ref{ymuresummation},\ref{ds-DeltamuResult}), and the equations
from Section 2.2 of Ref.~\cite{Gorbahn:2009pp}, respectively.
The $b$-quark mass $m_b$ used to calculate the Yukawa coupling is the
SM \MSbar\ mass with $5$ active quark flavours at the scale $M_Z$, obtained as described
in Ref.~\cite{Baer:2002ek}. Non-$\tan\beta$-enhanced matching
corrections to $m_b$ in the \DRbar\ scheme in the MSSM are
neglected since they would amount to negligible three-loop
contributions to $\amu^{\text{SUSY}}$.

The two-loop
corrections of type 2L(a) are implemented in the approximation of
photonic Barr-Zee diagrams~\cite{ArhribBaek,ChenGeng,HSW03,HSW04},
$\amuTLa=a_\mu^{(\chi\gamma H)}+a_\mu^{(\tilde{f}\gamma H)}$,
as provided by Ref.~\cite{DSreview}: 
\begin{subequations}
\begin{align}
\label{ds-amuchigammaphi}
\qquad
a_\mu^{(\chi\gamma H)} &= 
\frac{\alpha(M_Z)^2 m_\mu^2}{8\pi^2 M_W^2 s_W^2}\
\sum_{k=1,2}\Big[
{\rm Re}[ \lambda_\mu^{A^0} \lambda_{\chi^-_k}^{A^0}]
\ f_{PS}(m_{\chi^-_k}^2/M_{A^0}^2)
+
\sum_{S=h^0,H^0} {\rm Re}[\lambda_\mu^S \lambda_{\chi^-_k}^S ]
\ f_S(m_{\chi^-_k}^2/M_S^2)
\Big]
,\\
\label{ds-amusfgammaphi}
a_\mu^{(\tilde{f}\gamma H)}& =
\frac{\alpha(M_Z)^2 m_\mu^2}{8\pi^2 M_W^2 s_W^2}\
\sum_{\tilde{f}=\tilde{t},\tilde{b},\tilde{\tau}}\sum_{i=1,2}
\Big[
(N_c Q^2)_{\tilde{f}}
\sum_{S=h^0,H^0}
{\rm Re}[ \lambda_\mu^S \lambda_{\tilde{f}_i}^S]
\ f_{\tilde{f}}(m_{\tilde{f}_i}^2/M_S^2)
\Big]
\end{align}
\end{subequations}
with the Higgs coupling factors, which take into account the
$\tan\beta$-resummed Yukawa couplings of the muon, $\tau$-lepton and
$b$-quark, 
\begin{subequations}
\begin{align}
\lambda_\mu^{\{h^0,H^0,A^0\}} &= \frac{\sqrt2\,s_W M_W\,y_\mu }{m_\mu\, e}\left\{
-{s_\alpha},{c_\alpha},{s_\beta}
\right\},\\
%
\lambda_{\chi^-_k}^{\{h^0,H^0,A^0\}} &= \frac{\sqrt2 M_W}{m_{\chi^-_k}}
\big(U_{k1}V_{k2}\big\{ c_\alpha, s_\alpha, -c_\beta\big\}
+U_{k2}V_{k1}\big\{ -s_\alpha,c_\alpha,-s_\beta \big\}\big)
,\\
\lambda_{\tilde{t}_i}^{\{h^0,H^0\}} &=
\frac{2m_t}
{m_{\tilde{t}_i}^2 s_\beta}
\big(+\mu^*\big\{s_\alpha,-c_\alpha\big\}+A_t\big\{c_\alpha,s_\alpha\big\}\big)
\  (U^{\tilde{t}}_{i1})^*\  U^{\tilde{t}}_{i2} 
,\\
\lambda_{\tilde{b}_i}^{\{h^0,H^0\}} &=
\frac{2\sqrt2\,s_W M_W\,y_b}
{m_{\tilde{b}_i}^2 \, e}
\big(-\mu^*\big\{c_\alpha,s_\alpha\big\}+A_b\big\{-s_\alpha,c_\alpha\big\}\big)
\  (U^{\tilde{b}}_{i1})^* \   U^{\tilde{b}}_{i2} 
,\\
\lambda_{\tilde{\tau}_i}^{\{h^0,H^0\}} &=
\frac{2 \sqrt2\, s_WM_W\,y_\tau}
{m_{\tilde{\tau}_i}^2 \, e}
\big(-\mu^*\big\{c_\alpha,s_\alpha\big\}
+A_\tau\big\{-s_\alpha,c_\alpha\big\}\big)
\  (U^{\tilde{\tau}}_{i1})^* \   U^{\tilde{\tau}}_{i2} 
.
\end{align}
\end{subequations}
Here we have abbreviated $A_t=A^u_{3,3}$, $A_b=A^d_{3,3}$,
$A_\tau=A^e_{3,3}$ and $s_\gamma=\sin\gamma$, $c_\gamma=\cos\gamma$, $t_\gamma=\tan\gamma$. The loop functions are given as
\begin{subequations}
\begin{align}
f_S(z) &= (2z-1)f_{PS}(z) - 2z(2+\log z),\\
f_{\tilde{f}}(z) &=\frac{z}{2}\Big[2+\log z-f_{PS}(z)\Big]
,\\
f_{PS}(z) &= z\int_0^1 \frac{{\rm d}x\log\frac{x(1-x)}{z}}{x(1-x)-z}
=
\frac{2z}{y}\Big[{\rm Li}_2\Big(1-\frac{1-y}{2z}\Big)
                -{\rm Li}_2\Big(1-\frac{1+y}{2z}\Big) \Big]
\end{align}
\end{subequations}
with $y=\sqrt{1-4z}$. Note that $f_{PS}(z)$ is real and analytic even
for $z\ge1/4$.

\newcommand{\amuFSfL}{a_{\mu}^{{\rm 2L}, f{\tilde f}\, {\rm LL}}}

\newcommand{\muDR}{Q}
\newcommand{\amuWHnu}{a_{\mu}^{1{\rm L}}({\tilde W}\text{--}{\tilde H},{\tilde \nu}_{\mu})} 
\newcommand{\amuWHmuL}{a_{\mu}^{1{\rm L}}({\tilde W}\text{--}{\tilde H},{\tilde \mu}_{L})} 
\newcommand{\amuBHmuL}{a_{\mu}^{1{\rm L}}({\tilde B}\text{--}{\tilde H},{\tilde \mu}_{L})}
\newcommand{\amuBHmuR}{a_{\mu}^{1{\rm L}}({\tilde B}\text{--}{\tilde H},{\tilde \mu}_{R})}
\newcommand{\amuBmuLmuR}{a_{\mu}^{1{\rm L}}({\tilde B},{\tilde \mu}_{L}\text{--}{\tilde \mu}_{R})} 

\newcommand{\amuWHnutwoL}{a_{\mu}^{2{\rm L}, f{\tilde f}\, {\rm LL}}({\tilde W}\text{--}{\tilde H},{\tilde \nu}_{\mu})} 
\newcommand{\amuWHmuLtwoL}{a_{\mu}^{2{\rm L},f{\tilde f}\, {\rm LL}}({\tilde W}\text{--}{\tilde H},{\tilde \mu}_{L})} 
\newcommand{\amuBHmuLtwoL}{a_{\mu}^{2{\rm L},f{\tilde f}\, {\rm LL}}({\tilde B}\text{--}{\tilde H},{\tilde \mu}_{L})}
\newcommand{\amuBHmuRtwoL}{a_{\mu}^{2{\rm L},f{\tilde f}\, {\rm LL}}({\tilde B}\text{--}{\tilde H},{\tilde \mu}_{R})}
\newcommand{\amuBmuLmuRtwoL}{a_{\mu}^{2{\rm L},f{\tilde f}\, {\rm LL}}({\tilde B},{\tilde \mu}_{L}\text{--}{\tilde \mu}_{R})} 
\newcommand{\logscale}{m_{\text{SUSY}}}

\newcommand{\Deltagone}{\Delta_{g_{1}}}
\newcommand{\Deltagtwo}{\Delta_{g_{2}}}
\newcommand{\DeltaYukHiggsino}{\Delta_{\tilde H}}
\newcommand{\DeltaYukBinoHiggsino}{\Delta_{{\tilde B}{\tilde H}}}
\newcommand{\DeltaYukWinoHiggsino}{\Delta_{{\tilde W}{\tilde H}}}
\newcommand{\Deltatanbe}{\Delta_{t_{\beta}}}

\newcommand{\constA}{0.015}
\newcommand{\constB}{0.015}
\newcommand{\constC}{0.015}
\newcommand{\constD}{0.04}
\newcommand{\constE}{0.03}

The two-loop fermion/sfermion-loop contributions of
Refs.~\cite{Fargnoli:2013zda,Fargnoli:2013zia} are implemented in the
leading-logarithmic approximation $\amuFSfL$ given in
Ref.~\cite{Fargnoli:2013zia}:

\begin{align}
\begin{split}
\amuFSfL =&\
\amuWHnu\,
\Big(\Deltagtwo + \DeltaYukHiggsino + \DeltaYukWinoHiggsino + \Deltatanbe + \constA\Big)\\*
&+\amuWHmuL\,
\Big(\Deltagtwo + \DeltaYukHiggsino + \DeltaYukWinoHiggsino + \Deltatanbe + \constB\Big)\\*
&+\amuBHmuL\,
\Big(\Deltagone + \DeltaYukHiggsino + \DeltaYukBinoHiggsino + \Deltatanbe + \constC\Big)\\*
&+\amuBHmuR\,
\Big(\Deltagone + \DeltaYukHiggsino + \DeltaYukBinoHiggsino + \Deltatanbe + \constD\Big)\\*
&+\amuBmuLmuR\,
\Big(\Deltagone + \Deltatanbe + \constE\Big).
\label{eq:twolooplogapprox}
\end{split}
\end{align}
This is based on the mass-insertion approximation of the one-loop result
given in Ref.~\cite{Cho:2011rk}, and correction factors $\Delta_i$ as
well as numerical constants obtained via a fit to the exact
result. The mass-insertion approximation reads
\begin{subequations}
\label{eq:onelooplogapprox}
\begin{align}
\amuWHnu &= \frac{g_{2}^{2}}{8 \pi ^{2}} \frac{m_{\mu} ^{2} M_{2}}{m_{{\tilde \nu}_{\mu}}^{4}}\,\mu \tan\beta\, 
F_{a}\left(\frac{M_{2}^{2}}{m_{{\tilde \nu}_{\mu}}^{2}},\frac{\mu^{2}}{m_{{\tilde \nu}_{\mu}}^{2}}\right), \\* 
\amuWHmuL &= - \frac{g_{2}^{2}}{16 \pi ^{2}} \frac{m_{\mu}^{2} M_{2}}{M_{L 2}^{4}}\,\mu \tan\beta\,
F_{b}\left(\frac{M_{2}^{2}}{M_{L 2}^{2}},\frac{\mu^{2}}{M_{L 2}^{2}}\right),\\*
\amuBHmuL &= \frac{g_{1}^{2}}{16 \pi ^{2}}\frac{m_{\mu}^{2} M_{1}}{M_{L 2}^{4}}\,\mu \tan\beta\,
F_{b}\left(\frac{M_{1}^{2}}{M_{L 2}^{2}},\frac{\mu^{2}}{M_{L 2}^{2}}\right), \\* 
\amuBHmuR &= - \frac{g_{1}^{2}}{8 \pi ^{2}} \frac{m_{\mu}^{2} M_{1}}{M_{E 2}^{4}}\,\mu \tan\beta\, 
F_{b}\left(\frac{M_{1}^{2}}{M_{E 2}^{2}},\frac{\mu^{2}}{M_{E 2}^{2}}\right), \\* 
\amuBmuLmuR &= \frac{g_{1}^{2}}{8 \pi ^{2}} \frac{m_{\mu}^{2}}{M_{1}^{3}}\,\mu \tan\beta\, 
F_{b}\left(\frac{M_{L 2}^{2}}{M_{1}^{2}},\frac{M_{E 2}^{2}}{M_{1}^{2}}\right),
\end{align} 
\end{subequations}
 where
\begin{subequations}
\begin{align}
F_{a}(x, y) &= -\frac{G_{3}(x) - G_{3}(y)}{x - y}, \\
F_{b}(x, y) &= -\frac{G_{4}(x) - G_{4}(y)}{x - y},\\
G_{3} (x) &= \frac{1}{2(x-1)^{3}}\Big[ (x-1)(x-3) + 2\log x\Big],\\
G_{4} (x) &= \frac{1}{2(x-1)^{3}}\Big[ (x-1)(x+1) - 2 x \log x \Big] .
\end{align}
\end{subequations}
The shifts $\Delta_i$ are given by a slight generalization of the form
in Ref.~\cite{Fargnoli:2013zia}, dropping the assumption of
universality of the first two sfermion generations:
\begin{subequations}
\begin{align}
\begin{split}
  \Deltagone &=
  \begin{aligned}[t]
     \frac{g_{1}^{2}}{16 \pi ^2} \frac{4}{3}\sum_{i=1,2,3}\biggl(
     & \frac{4}{3} \log\frac{M_{U i}}{\logscale} + \frac{1}{3} \log\frac{M_{D i}}{\logscale}\\
     & + \frac{1}{6} \log\frac{M_{Q i}}{\logscale} + \log\frac{M_{Ei}}{\logscale} + \frac{1}{2} \log\frac{M_{L i}}{\logscale}\biggr),
   \end{aligned}
\end{split}\\
\Deltagtwo &= \frac{g_{2}^{2}}{16 \pi ^{2}} \frac{4}{3}\sum_{i=1,2,3}
\left( \frac{3}{2}\log\frac{M_{Q i}}{\logscale} 
+ \frac{1}{2}\log\frac{M_{L i}}{\logscale}
\right),\\*
\begin{split}
  \DeltaYukHiggsino &=
  \begin{aligned}[t]
    \frac{1}{16 \pi ^2} \frac{1}{2}\biggl(
    & 3 y_{t}^{2} \log\frac{M_{U 3}}{\logscale} + 3 y_{b}^{2} \log\frac{M_{D 3}}{\logscale} + 3 (y_{t}^{2} + y_{b}^{2}) \log\frac{M_{Q 3}}{\logscale}\\
    &+ y_{\tau}^{2} \log\frac{M _{E 3}}{\logscale} + y_{\tau}^{2} \log\frac{M _{L 3}}{\logscale}\biggr),
  \end{aligned}
\end{split}\\*
\DeltaYukBinoHiggsino &= \frac{1}{16 \pi ^{2}} 
y_{t}^{2} \left(2 \log\frac{M_{Q 3}}{\logscale} - 8 \log\frac{M_{U 3}}{\logscale}\right), \\
\DeltaYukWinoHiggsino &= \frac{1}{16 \pi ^{2}} 
y_{t}^{2} \left(-6\log\frac{M_{Q 3}}{\logscale} \right),\\ 
\Deltatanbe &= \frac{1}{16 \pi ^{2}}(3 y_{b}^{2} - 3 y_{t}^{2} + y_{\tau}^{2})\log\frac{\muDR}{\logscale}, 
\end{align}
\end{subequations}
where $\logscale = \min (\lvert\mu\rvert, \lvert M_{1}\rvert, \lvert
M_{2}\rvert, M_{L 2}, M_{E 2})$ and $\muDR$ is the renormalization
scale used to define $\tan\beta$. In order to apply the resummation of
$\tan\beta$-enhanced contributions, we employ
\begin{align}
\left[\amuFSfL\right]_{t_\beta\text{-resummed}}=\frac{\amuFSfL}{1+\Delta_\mu}.
\end{align}

\section{Example input file in SLHA format}
\label{sec:example-slha-file}

\begin{lstlisting}[caption={example SLHA input file \code{input/example.slha} with MSSM parameter point in the style of BM1~\cite{Fargnoli:2013zda}}, label=example-slha-file]
Block GM2CalcConfig
     0     3     # output format (0 = minimal, 1 = detailed,
                 #  2 = NMSSMTools, 3 = SPheno, 4 = GM2Calc)
     1     2     # loop order (0, 1 or 2)
     2     1     # disable/enable tan(beta) resummation (0 or 1)
     3     0     # force output (0 or 1)
     4     0     # verbose output (0 or 1)
     5     0     # calculate uncertainty
Block GM2CalcInput
     1     0.00775531       # alpha(MZ)             [1L]
     2     0.00729735       # alpha(0)              [2L]
Block SMINPUTS
     3     0.1184           # alpha_s(MZ) SM MSbar  [2L]
     4     9.11876000E+01   # MZ(pole)              [1L]
     5     4.18000000E+00   # mb(mb) SM MSbar       [2L]
     6     1.73340000E+02   # mtop(pole)            [2L]
     7     1.77700000E+00   # mtau(pole)            [2L]
     8     0.00000000E+00   # mnu3(pole)            [irrelevant]
    11     0.000510998928   # melectron(pole)       [irrelevant]
    12     0.00000000E+00   # mnu1(pole)            [irrelevant]
    13     0.1056583715     # mmuon(pole)           [1L]
    14     0.00000000E+00   # mnu2(pole)            [irrelevant]
    21     4.76052706E-03   # md                    [irrelevant]
    22     2.40534062E-03   # mu                    [irrelevant]
    23     1.04230487E-01   # ms                    [irrelevant]
    24     1.27183378E+00   # mc                    [irrelevant]
Block MASS                      # Mass spectrum
        24     8.03773317e+01   # W                 [1L]
        25     1.25712136e+02   # h0                [irrelevant]
        35     1.50002058e+03   # H0                [irrelevant]
        36     1.50000000e+03   # A0                [2L]
        37     1.50241108e+03   # H+                [irrelevant]
   1000021     2.27707784e+03   # ~g                [irrelevant]
   1000022     2.01611468e+02   # neutralino(1)     [1L]
   1000023     4.10040273e+02   # neutralino(2)     [1L]
   1000024     4.09989890e+02   # chargino(1)       [1L]
   1000025    -5.16529941e+02   # neutralino(3)     [1L]
   1000035     5.45628749e+02   # neutralino(4)     [1L]
   1000037     5.46057190e+02   # chargino(2)       [1L]
   1000001     7.06303219e+03   # ~d_L              [irrelevant]
   1000002     7.06271372e+03   # ~u_L              [irrelevant]
   1000003     7.06305914e+03   # ~s_L              [irrelevant]
   1000004     7.06274067e+03   # ~c_L              [irrelevant]
   1000005     7.10826650e+03   # ~b_1              [irrelevant]
   1000006     7.16237728e+03   # ~t_1              [irrelevant]
   1000011     5.25167746e+02   # ~e_L              [irrelevant]
   1000012     5.18841083e+02   # ~nu_e_L           [irrelevant]
   1000013     5.25187016e+02   # ~mu_L             [1L]
   1000014     5.18860573e+02   # ~nu_mu_L          [1L]
   1000015     3.00151415e+03   # ~tau_1            [irrelevant]
   1000016     3.00880751e+03   # ~nu_tau_L         [irrelevant]
   2000001     7.04938393e+03   # ~d_R              [irrelevant]
   2000002     7.05136356e+03   # ~u_R              [irrelevant]
   2000003     7.04943468e+03   # ~s_R              [irrelevant]
   2000004     7.05136702e+03   # ~c_R              [irrelevant]
   2000005     7.16151312e+03   # ~b_2              [irrelevant]
   2000006     7.19205727e+03   # ~t_2              [irrelevant]
   2000011     5.05054724e+02   # ~e_R              [irrelevant]
   2000013     5.05095249e+02   # ~mu_R             [1L]
   2000015     3.01426808e+03   # ~tau_2            [irrelevant]
Block HMIX Q= 1.00000000e+03
     1     4.89499929e+02      # mu(Q) MSSM DRbar   [initial guess]
     2     3.93371545e+01      # tan(beta)(Q) MSSM DRbar Feynman gauge [1L]
Block MSOFT Q= 1.00000000e+03  # MSSM DRbar SUSY breaking parameters
     1     2.00000000e+02      # M_1(Q)             [initial guess]
     2     4.00000000e+02      # M_2(Q)             [initial guess]
     3     2.00000000e+03      # M_3(Q)             [2L]
    31     5.00000000e+02      # meL(Q)             [2L]
    32     5.00000000e+02      # mmuL(Q)            [irrelevant]
    33     3.00000000e+03      # mtauL(Q)           [2L]
    34     4.99999999e+02      # meR(Q)             [2L]
    35     4.99999999e+02      # mmuR(Q)            [initial guess]
    36     3.00000000e+03      # mtauR(Q)           [2L]
    41     7.00000000e+03      # mqL1(Q)            [2L]
    42     7.00000000e+03      # mqL2(Q)            [2L]
    43     6.99999999e+03      # mqL3(Q)            [2L]
    44     7.00000000e+03      # muR(Q)             [2L]
    45     7.00000000e+03      # mcR(Q)             [2L]
    46     6.99999999e+03      # mtR(Q)             [2L]
    47     7.00000000e+03      # mdR(Q)             [2L]
    48     7.00000000e+03      # msR(Q)             [2L]
    49     7.00000000e+03      # mbR(Q)             [2L]
Block AU Q= 1.00000000e+03
  3  3     1.57871614e-05      # At(Q)              [2L]
Block AD Q= 1.00000000e+03
  3  3     8.99561673e-06      # Ab(Q)              [2L]
Block AE Q= 1.00000000e+03
  2  2     2.84230475e-06      # Amu(Q) MSSM DR-bar [1L]
  3  3     3.02719242e-06      # Atau(Q)            [2L]
\end{lstlisting}

\section{Example input file in \OURPROGRAM-specific format}
\label{sec:example-gm2calc-file}

\begin{lstlisting}[caption={example GM2Calc input file \code{input/example.gm2}},label=gm2calc-slha-input-example]
Block GM2CalcConfig
     0     1     # output format (0 = minimal, 1 = detailed,
                 #  2 = NMSSMTools, 3 = SPheno, 4 = GM2Calc)
     1     2     # loop order (0, 1 or 2)
     2     1     # disable/enable tan(beta) resummation (0 or 1)
     3     0     # force output (0 or 1)
     4     0     # verbose output (0 or 1)
     5     0     # calculate uncertainty
Block GM2CalcInput
     0     8.66360379E+02   # ren. scale Q          [2L]
     1     0.00775531       # alpha(MZ)             [1L]
     2     0.00729735       # alpha(0)              [2L]
     3     10               # tan(beta) DR-bar at Q [1L]
     4     619.858          # Mu parameter on-shell [1L]
     5     211.722          # M1 on-shell           [1L]
     6     401.057          # M2 on-shell           [1L]
     7     1.10300877E+03   # M3                    [2L]
     8     707.025          # MA(pole)              [2L]
     9     3.51653258E+02   # msl(1,1)              [2L]
    10     356.09           # msl(2,2) on-shell     [1L]
    11     3.50674223E+02   # msl(3,3)              [2L]
    12     2.21215037E+02   # mse(1,1)              [2L]
    13     225.076          # mse(2,2) on-shell     [1L]
    14     2.18022142E+02   # mse(3,3)              [2L]
    15     1.00711403E+03   # msq(1,1)              [2L]
    16     1.00711149E+03   # msq(2,2)              [2L]
    17     9.29083096E+02   # msq(3,3)              [2L]
    18     9.69369660E+02   # msu(1,1)              [2L]
    19     9.69366965E+02   # msu(2,2)              [2L]
    20     7.99712943E+02   # msu(3,3)              [2L]
    21     9.64756473E+02   # msd(1,1)              [2L]
    22     9.64753818E+02   # msd(2,2)              [2L]
    23     9.60016201E+02   # msd(3,3)              [2L]
    24     0                # Ae(1,1)               [irrelevant]
    25     -2.93720212E+02  # Ae(2,2) DR-bar        [1L]
    26     -2.92154796E+02  # Ae(3,3)               [2L]
    27     0                # Ad(1,1)               [irrelevant]
    28     0                # Ad(2,2)               [irrelevant]
    29     -1.28330100E+03  # Ad(3,3)               [2L]
    30     0                # Au(1,1)               [irrelevant]
    31     0                # Au(2,2)               [irrelevant]
    32     -8.70714986E+02  # Au(3,3)               [2L]
Block SMINPUTS
     3     0.1184           # alpha_s(MZ) SM MS-bar [2L]
     4     91.1876          # M_Z(pole)             [1L]
     5     4.18             # m_b(m_b) SM MS-bar    [2L]
     6     173.34           # M_top(pole)           [2L]
     7     1.777            # M_tau(pole)           [2L]
     8     0.00000000E+00   # mnu3(pole)            [irrelevant]
     9     80.385           # M_W(pole)             [1L]
    11     0.000510998928   # melectron(pole)       [irrelevant]
    12     0.00000000E+00   # mnu1(pole)            [irrelevant]
    13     0.1056583715     # M_muon(pole)          [1L]
    14     0.00000000E+00   # mnu2(pole)            [irrelevant]
    21     4.76052706E-03   # md                    [irrelevant]
    22     2.40534062E-03   # mu                    [irrelevant]
    23     1.04230487E-01   # ms                    [irrelevant]
    24     1.27183378E+00   # mc                    [irrelevant]
\end{lstlisting}

\end{appendix}


\clearpage
\begin{thebibliography}{999}
%

\bibitem{Bennett:2006fi} G.W. Bennett, et al.,
(Muon $(g-2)$ Collaboration), Phys. Rev. D {\bf 73}, 072003 (2006).

\bibitem{Davier}
  M.~Davier, A.~Hoecker, B.~Malaescu and Z.~Zhang,
  Eur.\ Phys.\ J.\ C {\bf 71} (2011) 1515
   [Erratum-ibid.\ C {\bf 72} (2012) 1874]
  [arXiv:1010.4180 [hep-ph]].

\bibitem{HMNT}
  K.~Hagiwara, R.~Liao, A.~D.~Martin, D.~Nomura and T.~Teubner,
  J.\ Phys.\ G {\bf 38} (2011) 085003
  [arXiv:1105.3149 [hep-ph]].

\bibitem{Kinoshita2012}
  T.~Aoyama, M.~Hayakawa, T.~Kinoshita and M.~Nio,
  Phys.\ Rev.\ Lett.\  {\bf 109} (2012) 111808
  [arXiv:1205.5370 [hep-ph]].

\bibitem{Gnendiger:2013pva}
  C.~Gnendiger, D.~St{\"o}ckinger and H.~St{\"o}ckinger-Kim,
  Phys.\ Rev.\ D {\bf 88} (2013) 053005
  [arXiv:1306.5546 [hep-ph]].

\bibitem{Kataev:2012kn}
  A.~L.~Kataev,
  Phys.\ Rev.\ D {\bf 86} (2012) 013010
  [arXiv:1205.6191 [hep-ph]].
\bibitem{SteinhauserQED}
  R.~Lee, P.~Marquard, A.~V.~Smirnov, V.~A.~Smirnov and M.~Steinhauser,
  JHEP {\bf 1303} (2013) 162
  [arXiv:1301.6481 [hep-ph]];
  A.~Kurz, T.~Liu, P.~Marquard and M.~Steinhauser,
  Nucl.\ Phys.\ B {\bf 879} (2014) 1
  [arXiv:1311.2471 [hep-ph]];
  A.~Kurz, T.~Liu, P.~Marquard, A.~V.~Smirnov, V.~A.~Smirnov and M.~Steinhauser,
  arXiv:1508.00901 [hep-ph].




\bibitem{JegerlehnerSzafron}
  F.~Jegerlehner and R.~Szafron,
  Eur.\ Phys.\ J.\ C {\bf 71} (2011) 1632
  [arXiv:1101.2872 [hep-ph]].

\bibitem{Benayoun:2012wc}
  M.~Benayoun, P.~David, L.~DelBuono and F.~Jegerlehner,
  Eur.\ Phys.\ J.\ C {\bf 73} (2013) 2453
  [arXiv:1210.7184 [hep-ph]];
  arXiv:1507.02943 [hep-ph].

\bibitem{Kurz:2014wya}
  A.~Kurz, T.~Liu, P.~Marquard and M.~Steinhauser,
  Phys.\ Lett.\ B {\bf 734} (2014) 144
  [arXiv:1403.6400 [hep-ph]].

\bibitem{Colangelo:2014qya}
  G.~Colangelo, M.~Hoferichter, A.~Nyffeler, M.~Passera and P.~Stoffer,
  Phys.\ Lett.\ B {\bf 735} (2014) 90
  [arXiv:1403.7512 [hep-ph]].


\bibitem{Colangelo:2014}
  G.~Colangelo, M.~Hoferichter, M.~Procura and P.~Stoffer,
  JHEP {\bf 1409} (2014) 091
  [arXiv:1402.7081 [hep-ph]];
  G.~Colangelo, M.~Hoferichter, B.~Kubis, M.~Procura and P.~Stoffer,
  Phys.\ Lett.\ B {\bf 738} (2014) 6
  [arXiv:1408.2517 [hep-ph]];
  G.~Colangelo, M.~Hoferichter, M.~Procura and P.~Stoffer,
  JHEP {\bf 1509} (2015) 074
  [arXiv:1506.01386 [hep-ph]].

\bibitem{Pauk:2014rfa}
  V.~Pauk and M.~Vanderhaeghen,
  Phys.\ Rev.\ D {\bf 90} (2014) 11,  113012
  [arXiv:1409.0819 [hep-ph]].


\bibitem{Blum}
  T.~Blum, S.~Chowdhury, M.~Hayakawa and T.~Izubuchi,
  Phys.\ Rev.\ Lett.\  {\bf 114} (2015) 1,  012001
  [arXiv:1407.2923 [hep-lat]].
%
  L.~Jin, T.~Blum, N.~Christ, M.~Hayakawa, T.~Izubuchi and C.~Lehner,
  arXiv:1509.08372 [hep-lat].

\bibitem{Ablikim:2015orh}
  M.~Ablikim {\it et al.} [BESIII Collaboration],
  arXiv:1507.08188 [hep-ex].




\bibitem{JegerlehnerNyffeler}
  F.~Jegerlehner and A.~Nyffeler,
  Phys.\ Rept.\  {\bf 477} (2009) 1
  [arXiv:0902.3360 [hep-ph]].
\bibitem{Miller:2012opa}
  J.~P.~Miller, E.~de Rafael, B.~L.~Roberts and D.~St\"ockinger,
  Ann.\ Rev.\ Nucl.\ Part.\ Sci.\  {\bf 62} (2012) 237.
\bibitem{Blum:2013xva}
  T.~Blum, A.~Denig, I.~Logashenko, E.~de Rafael, B.~L.~Roberts, T.~Teubner and G.~Venanzoni,
  arXiv:1311.2198 [hep-ph].
\bibitem{Benayoun:2014tra}
  M.~Benayoun, J.~Bijnens, T.~Blum, I.~Caprini, G.~Colangelo, H.~Czyz, A.~Denig and C.~A.~Dominguez {\it et al.},
  arXiv:1407.4021 [hep-ph].












\bibitem{Carey:2009zzb}
  R.~M.~Carey, K.~R.~Lynch, J.~P.~Miller, B.~L.~Roberts, W.~M.~Morse, Y.~K.~Semertzides, V.~P.~Druzhinin and B.~I.~Khazin {\it et al.},
  FERMILAB-PROPOSAL-0989;
  B.~L.~Roberts,
  Chin.\ Phys.\ C {\bf 34} (2010) 741
  [arXiv:1001.2898 [hep-ex]];
  J.~Grange {\it et al.}  [Muon g-2 Collaboration],
  arXiv:1501.06858 [physics.ins-det].
\bibitem{Iinuma:2011zz}
  H.~Iinuma [J-PARC New g-2/EDM experiment Collaboration],
  J.\ Phys.\ Conf.\ Ser.\  {\bf 295} (2011) 012032.



\bibitem{WhitePaper} D.~W.~Hertzog, J.~P.~Miller, E.~de Rafael, B.~Lee
  Roberts and D.~St\"ockinger, 
arXiv:0705.4617.  



\bibitem{CzM}
  A.~Czarnecki and W.~J.~Marciano,
  Phys.\ Rev.\ D {\bf 64} (2001) 013014
  [hep-ph/0102122].


\bibitem{MartinWells}
                   S.~Martin and J.~Wells,
                   {Phys. Rev. D} {\bf 64} (2001) 035003.  

\bibitem{DSreview} D.~St\"ockinger, 
J.\ Phys.\ G {\bf 34} (2007) R45. 



\bibitem{Cho:2011rk}
  G.~-C.~Cho, K.~Hagiwara, Y.~Matsumoto and D.~Nomura,
  JHEP {\bf 1111} (2011) 068
  [arXiv:1104.1769 [hep-ph]].









\bibitem{Endo}
  M.~Endo, K.~Hamaguchi, S.~Iwamoto and N.~Yokozaki,
  Phys.\ Rev.\ D {\bf 84} (2011) 075017
  [arXiv:1108.3071 [hep-ph]].
%
  M.~Endo, K.~Hamaguchi, S.~Iwamoto, K.~Nakayama and N.~Yokozaki,
  Phys.\ Rev.\ D {\bf 85} (2012) 095006
  [arXiv:1112.6412 [hep-ph]].
  M.~Endo, K.~Hamaguchi, S.~Iwamoto and N.~Yokozaki,
  Phys.\ Rev.\ D {\bf 85} (2012) 095012
  [arXiv:1112.5653 [hep-ph]].
  M.~Endo, K.~Hamaguchi, S.~Iwamoto and T.~Yoshinaga,
  JHEP {\bf 1401} (2014) 123
  [arXiv:1303.4256 [hep-ph]].
%
  M.~Endo, K.~Hamaguchi, T.~Kitahara and T.~Yoshinaga,
  JHEP {\bf 1311} (2013) 013
  [arXiv:1309.3065 [hep-ph]].
%
  M.~Endo, K.~Hamaguchi, S.~Iwamoto, T.~Kitahara and T.~Moroi,
  Phys.\ Lett.\ B {\bf 728} (2014) 274
  [arXiv:1310.4496 [hep-ph]].

\bibitem{Evans:2012hg}
  J.~L.~Evans, M.~Ibe, S.~Shirai and T.~T.~Yanagida,
  Phys.\ Rev.\ D {\bf 85} (2012) 095004
  [arXiv:1201.2611 [hep-ph]].



\bibitem{Ibe:2013oha}
  M.~Ibe, T.~T.~Yanagida and N.~Yokozaki,
  JHEP {\bf 1308} (2013) 067
  [arXiv:1303.6995 [hep-ph]].

\bibitem{Akula:2013ioa}
  S.~Akula and P.~Nath,
  Phys.\  Rev.\  D 87, {\bf 115022} (2013)
  [arXiv:1304.5526 [hep-ph]].




\bibitem{Bhattacharyya:2013xma}
  G.~Bhattacharyya, B.~Bhattacherjee, T.~T.~Yanagida and N.~Yokozaki,
  Phys.\ Lett.\ B {\bf 730} (2014) 231
  [arXiv:1311.1906 [hep-ph]].
\bibitem{Mohanty:2013soa}
  S.~Mohanty, S.~Rao and D.~P.~Roy,
  JHEP {\bf 1309} (2013) 027
  [arXiv:1303.5830 [hep-ph]].
\bibitem{Gogoladze:2014cha}
  I.~Gogoladze, F.~Nasir, Q.~Shafi and C.~S.~Un,
  Phys.\ Rev.\ D {\bf 90} (2014) 3, 035008
  [arXiv:1403.2337 [hep-ph]].

\bibitem{Kersten:2014xaa}
  J.~Kersten, J.-h.~Park, D.~St\"ockinger and L.~Velasco-Sevilla,
  JHEP {\bf 1408} (2014) 118
  [arXiv:1405.2972 [hep-ph]].
\bibitem{Chiu:2014oma}
  W.~C.~Chiu, C.~Q.~Geng and D.~Huang,
  Phys.\ Rev.\ D {\bf 91} (2015) 1,  013006
  [arXiv:1409.4198 [hep-ph]].

\bibitem{Badziak:2014kea}
  M.~Badziak, Z.~Lalak, M.~Lewicki, M.~Olechowski and S.~Pokorski,
  JHEP {\bf 1503} (2015) 003
  [arXiv:1411.1450 [hep-ph]].
\bibitem{Kowalska:2015zja}
  K.~Kowalska, L.~Roszkowski, E.~M.~Sessolo and A.~J.~Williams,
  arXiv:1503.08219 [hep-ph].
\bibitem{Wang:2015rli}
  F.~Wang, W.~Wang and J.~M.~Yang,
  arXiv:1504.00505 [hep-ph].


\bibitem{Calibbi:2015kja}
  L.~Calibbi, I.~Galon, A.~Masiero, P.~Paradisi and Y.~Shadmi,
  arXiv:1502.07753 [hep-ph].

\bibitem{Okada:2013ija} 
  N.~Okada, S.~Raza and Q.~Shafi,
  Phys.\ Rev.\ D {\bf 90}, no. 1, 015020 (2014)
  [arXiv:1307.0461 [hep-ph]].
  
\bibitem{Li:2014dna} 
  T.~Li and S.~Raza,
  Phys.\ Rev.\ D {\bf 91}, no. 5, 055016 (2015)
  [arXiv:1409.3930 [hep-ph]].

\bibitem{Allanach:2001kg} 
  B.~C.~Allanach,
  Comput.\ Phys.\ Commun.\  {\bf 143}, 305 (2002)
  [hep-ph/0104145].

\bibitem{spheno}
  W.~Porod,
  Comput.\ Phys.\ Commun.\  {\bf 153} (2003) 275
  [hep-ph/0301101].
  W.~Porod and F.~Staub,
  Comput.\ Phys.\ Commun.\  {\bf 183} (2012) 2458
  [arXiv:1104.1573 [hep-ph]].


\bibitem{Djouadi:2002ze} 
  A.~Djouadi, J.~-L.~Kneur and G.~Moultaka,
  Comput.\ Phys.\ Commun.\  {\bf 176}, 426 (2007)
  [hep-ph/0211331].

\bibitem{Baer:1993ae} 
  H.~Baer, F.~E.~Paige, S.~D.~Protopopescu and X.~Tata,
  hep-ph/9305342.


\bibitem{Chowdhury:2011zr}
  D.~Chowdhury, R.~Garani and S.~K.~Vempati,
  Comput.\ Phys.\ Commun.\  {\bf 184} (2013) 899
  [arXiv:1109.3551 [hep-ph]].



\bibitem{Sarah} 
  F.~Staub,
  Comput.\ Phys.\ Commun.\  {\bf 181}, 1077 (2010)
  [arXiv:0909.2863 [hep-ph]].
%
  Comput.\ Phys.\ Commun.\  {\bf 182}, 808 (2011)
  [arXiv:1002.0840 [hep-ph]].
%
  Computer Physics Communications {\bf 184}, pp. 1792 (2013)
  [Comput.\ Phys.\ Commun.\  {\bf 184}, 1792 (2013)]
  [arXiv:1207.0906 [hep-ph]].
%
  Comput.\ Phys.\ Commun.\  {\bf 185} (2014) 1773
  [arXiv:1309.7223 [hep-ph]].



\bibitem{Athron:2014yba}
  P.~Athron, J.-h.~Park, D.~St\"{o}ckinger and A.~Voigt,
  Comput.\ Phys.\ Commun.\  {\bf 190} (2015) 139
  [arXiv:1406.2319 [hep-ph]].



\bibitem{Mahmoudi:2008tp}
  F.~Mahmoudi,
  Comput.\ Phys.\ Commun.\  {\bf 180} (2009) 1579
  [arXiv:0808.3144 [hep-ph]].

\bibitem{Hahn:2009zz}
  T.~Hahn, S.~Heinemeyer, W.~Hollik, H.~Rzehak and G.~Weiglein,
  Comput.\ Phys.\ Commun.\  {\bf 180} (2009) 1426.

\bibitem{Rosiek:2010ug}
  J.~Rosiek, P.~Chankowski, A.~Dedes, S.~Jager and P.~Tanedo,
  Comput.\ Phys.\ Commun.\  {\bf 181} (2010) 2180
  [arXiv:1003.4260 [hep-ph]].

\bibitem{Lee:2012wa}
  J.~S.~Lee, M.~Carena, J.~Ellis, A.~Pilaftsis and C.~E.~M.~Wagner,
  Comput.\ Phys.\ Commun.\  {\bf 184} (2013) 1220
  [arXiv:1208.2212 [hep-ph]].


\bibitem{Skands:2003cj}
  P.~Z.~Skands {\it et al.},
  JHEP {\bf 0407} (2004) 036
  [hep-ph/0311123].




\bibitem{Fargnoli:2013zda}
  H.~G.~Fargnoli, C.~Gnendiger, S.~Pa\ss{}ehr, D.~St\"ockinger and H.~St\"ockinger-Kim,
  Phys.\ Lett.\ B {\bf 726} (2013) 717
  [arXiv:1309.0980 [hep-ph]].
\bibitem{Fargnoli:2013zia}
  H.~G.~Fargnoli, C.~Gnendiger, S.~Pa\ss{}ehr, D.~St\"ockinger and H.~St\"ockinger-Kim,
  JHEP {\bf 1402} (2014) 070
  [arXiv:1311.1775 [hep-ph]].

\bibitem{Marchetti:2008hw}
  S.~Marchetti, S.~Mertens, U.~Nierste and D.~St\"ockinger,
  Phys.\ Rev.\  D {\bf 79}, 013010 (2009)
  [arXiv:0808.1530 [hep-ph]].


\bibitem{Bach:2015doa}
  M.~Bach, J.-h.~Park, D.~St{\"o}ckinger and H.~St{\"o}ckinger-Kim,
  JHEP {\bf 1510} (2015) 026
  [arXiv:1504.05500 [hep-ph]].


\bibitem{moroi}
T.~Moroi,  
                     {Phys. Rev. D} {\bf  53} (1996) 6565
                     [Erratum-ibid.\ {\bf 56} (1997) 4424].



\bibitem{HSW03}
  S.~Heinemeyer, D.~St\"ockinger and G.~Weiglein,
  Nucl.\ Phys.\ B {\bf 690} (2004) 62.
%
\bibitem{HSW04}
  S.~Heinemeyer, D.~St\"ockinger and G.~Weiglein,
  Nucl.\ Phys.\ B {\bf 699} (2004) 103.

\bibitem{ArhribBaek}
  A.~Arhrib and S.~Baek,
  Phys.\ Rev.\ D {\bf 65} (2002) 075002
  [hep-ph/0104225].
  
\bibitem{ChenGeng}
  C.~H.~Chen and C.~Q.~Geng,
  Phys.\ Lett.\ B {\bf 511} (2001) 77
  [arXiv:hep-ph/0104151].


\bibitem{vonWeitershausen:2010zr}
  P.~von Weitershausen, M.~Sch\"afer, H.~St\"ockinger-Kim and D.~St\"ockinger,
  Phys.\ Rev.\ D {\bf 81} (2010) 093004
  [arXiv:1003.5820 [hep-ph]].


\bibitem{DG98}
  G.~Degrassi and G.~F.~Giudice,
  Phys.\ Rev.\ D {\bf 58} (1998) 053007
  [arXiv:hep-ph/9803384].



\bibitem{Dobrescu:2010mk}
  B.~A.~Dobrescu and P.~J.~Fox,
  Eur.\ Phys.\ J.\ C {\bf 70} (2010) 263
  [arXiv:1001.3147 [hep-ph]].
\bibitem{Altmannshofer:2010zt}
  W.~Altmannshofer and D.~M.~Straub,
  JHEP {\bf 1009} (2010) 078
  [arXiv:1004.1993 [hep-ph]].




\bibitem{HKRRSS}
  W.~Hollik, E.~Kraus, M.~Roth, C.~Rupp, K.~Sibold and D.~St\"ockinger,
  Nucl.\ Phys.\  B {\bf 639} (2002) 3.

\bibitem{tf}
  T.~Fritzsche and W.~Hollik,
  Eur.\ Phys.\ J.\  C {\bf 24}, 619 (2002).

\bibitem{Heidi}
  W.~Hollik and H.~Rzehak,
  Eur.\ Phys.\ J.\  C {\bf 32} (2003) 127.

\bibitem{Heinemeyer:2010mm}
  S.~Heinemeyer, H.~Rzehak and C.~Schappacher,
  Phys.\ Rev.\ D {\bf 82} (2010) 075010
  [arXiv:1007.0689 [hep-ph]].

\bibitem{Fritzsche:2011nr}
  T.~Fritzsche, S.~Heinemeyer, H.~Rzehak and C.~Schappacher,
  Phys.\ Rev.\ D {\bf 86} (2012) 035014
  [arXiv:1111.7289 [hep-ph]].

\bibitem{HeinemeyerNew}
  T.~Fritzsche, T.~Hahn, S.~Heinemeyer, H.~Rzehak and C.~Schappacher,
  [arXiv:1309.1692 [hep-ph]].

\bibitem{PDG}
  K.~A.~Olive {\it et al.} [Particle Data Group Collaboration],
  Chin.\ Phys.\ C {\bf 38} (2014) 090001.


\bibitem{Freitas:2002um}
  A.~Freitas and D.~St\"ockinger,
  Phys.\ Rev.\ D {\bf 66} (2002) 095014
  [hep-ph/0205281].

\bibitem{AbdusSalam:2011fc}
  S.~S.~AbdusSalam {\it et al.},
  Eur.\ Phys.\ J.\ C {\bf 71} (2011) 1835
  [arXiv:1109.3859 [hep-ph]].

\bibitem{Gorbahn:2009pp}
  M.~Gorbahn, S.~J\"ager, U.~Nierste and S.~Trine,
  Phys.\ Rev.\ D {\bf 84} (2011) 034030
  [arXiv:0901.2065 [hep-ph]].

\bibitem{Baer:2002ek}
H.~Baer, J.~Ferrandis, K.~ Melnikov and X.~Tata,
Phys.\ Rev.\ D {\bf 66} (2002) 074007,
[hep-ph/0207126].

\bibitem{Allanach:2008qq}
  B.~C.~Allanach {\it et al.},
  Comput.\ Phys.\ Commun.\  {\bf 180} (2009) 8
  [arXiv:0801.0045 [hep-ph]].








\end{thebibliography}
\end{document}